\documentclass[prx, preprint,aps,amsmath,amssymb,superscriptaddress,nofootinbib,floatfix]{revtex4-2}

\usepackage{braket}

\usepackage{graphicx}
\usepackage{dcolumn}
\usepackage{multirow}
\usepackage{url}
\usepackage{slashed}
\usepackage[colorlinks,linkcolor=black,anchorcolor=blue,citecolor=blue,urlcolor=blue]{hyperref}
\def\ccb#1{{\color{black} #1}}
\def\ccr#1{{\color{black} #1}}
\makeatletter
\patchcmd{\frontmatter@abstract@produce}
  {\vskip200\p@\@plus1fil
   \penalty-200\relax
   \vskip-200\p@\@plus-1fil}
  {}
  {}
  {}
\makeatother

\begin{document}

\title{A New Source of Phase Transition Gravitational Waves: Heavy Particle Braking across Bubble Walls}

\author{Dayun Qiu}

\author{Siyu Jiang}

\author{Fa Peng Huang}

\email{Corresponding Author.  huangfp8@sysu.edu.cn}

\affiliation{MOE Key Laboratory of TianQin Mission, TianQin Research Center for Gravitational Physics \& School of Physics and Astronomy, Frontiers Science Center for TianQin, Gravitational Wave Research Center of CNSA, Sun Yat-sen University (Zhuhai Campus), Zhuhai 519082, China}

\begin{abstract}
Motivated by the new heavy dark matter production mechanism from cosmic phase transition, we propose a novel mechanism for the generation of microscopic gravitational waves (GWs) during cosmological first-order phase transitions arising from the braking of heavy particles as they traverse bubble walls. Unlike the well-known sources such as bubble collisions, sound waves, or turbulence in the plasma, this mechanism originates from the direct interaction between massive particles and the expanding bubble wall. We use quantum field theory to rigorously compute the gravitational radiation.
The resulting GW spectrum exhibits distinctive features: The peak frequency is tightly correlated with the bubble wall velocity, while the peak amplitude scales as the fourth power of the heavy particle mass. These unique dependencies offer a new observational handle on particle physics beyond the Standard Model. 
We illustrate this mechanism within a specific model framework and demonstrate its viability. 
Our findings enrich the landscape of phase transition GW sources and open new avenues for more directly probing heavy particle dynamics and new physics models in the early universe.
\end{abstract}

\maketitle

\section{Introduction}

Gravitational waves (GWs) provide an unparalleled probe into the dynamics of the early universe, especially those stemming from cosmological first-order phase transitions. Such phase transitions are not only potential sources of observable GWs but also offer compelling frameworks for addressing fundamental cosmic puzzles, including the generation of dark matter and the observed baryon asymmetry. For instance, strong first-order phase transitions in the early universe, through the filtering effect of bubble walls on massive dark matter particles, can significantly suppress the relic abundance of dark matter while avoiding the unitarity limit~\cite{Baker:2019ndr,Chway:2019kft,Jiang:2023nkj}. Similarly, under conditions of primordial matter-antimatter asymmetry, the bubble wall's filtering effect can create an environment conducive to the formation of non-topological soliton dark matter~\cite{Witten:1984rs,Krylov:2013qe,Huang:2017kzu,Ponton:2019hux,Hong:2020est,Jiang:2023qbm,Jiang:2024zrb}. 

Moreover, due to the characteristic feature of strong phase transitions—namely, a large vacuum expectation value over temperature ratio $v_{\phi}/T \gg 1$—heavy particles such as right-handed neutrinos can decouple from the thermal bath immediately after the transition, thereby mitigating washout effects and preserving the generated asymmetry~\cite{Huang:2022vkf,Chun:2023ezg}.

In supercooled phase transitions, bubble walls often reach ultra-relativistic velocities. In such regimes, light particles traversing the bubble wall can produce superheavy particles via $1 \rightarrow 1$ or $1 \rightarrow 2$ processes. These superheavy particles may serve as dark matter candidates~\cite{Azatov:2021ifm,Azatov:2024crd,Ai:2024ikj} or as the origin of baryon asymmetry~\cite{Azatov:2021irb,Baldes:2021vyz}.

For these mechanisms of dark matter or matter-antimatter asymmetry generation, the most critical parameter is the expansion speed of the bubble wall. The bubble wall velocity for strong first-order phase transitions can be determined by solving the transport equations for particles crossing the bubble wall~\cite{Moore:1995ua,Moore:1995si,John:2000zq,Konstandin:2014zta,Kozaczuk:2015owa,Wang:2020zlf,Friedlander:2020tnq,Laurent:2020gpg,Cline:2021iff,Lewicki:2021pgr,Laurent:2022jrs,Jiang:2022btc,DeCurtis:2022hlx,Ekstedt:2024fyq} or approximated by assuming local thermal equilibrium~\cite{BarrosoMancha:2020fay,Balaji:2020yrx,Ai:2021kak,Wang:2022txy,Ai:2023see,Ai:2024btx,si2025bubblewallvelocitylocal}. For supercooling phase transitions, next-to-leading order (NLO) contributions are more significant, where particles can radiate vector bosons while crossing the bubble wall, exchanging momentum with it. Calculations for these bubble wall velocities are still under active investigation~\cite{Bodeker:2009qy,Bodeker:2017cim,Azatov:2020ufh,Hoche:2020ysm,Gouttenoire:2021kjv,Azatov:2023xem}.

Traditionally, GW signals from first-order phase transitions have been primarily attributed to macroscopic processes within the primordial plasma, such as bubble collisions, sound waves, and turbulence~\cite{Witten:1984rs,Kosowsky:1991ua,Kosowsky:1992vn,Kamionkowski:1993fg,Hindmarsh:2013xza}. These GWs are anticipated to be detectable by future GW observatories like LISA \cite{LISA:2017pwj}, TianQin \cite{TianQin:2015yph}, Taiji \cite{Hu:2017mde}, BBO \cite{Crowder:2005nr,Corbin:2005ny}, Cosmic Explorer (CE) \cite{Reitze:2019iox}, and Einstein Telescope (ET)~\cite{Sathyaprakash:2012jk,ET:2019dnz}. There are also some interesting proposals for the detection of high-frequency GW, like resonant cavities~\cite{Herman:2022fau} and plasma haloscopes~\cite{Capdevilla:2024cby}.
However, it is important to recognize that microscopic processes can also contribute to the GW background in a significant and model-independent manner. 
Since gravitational interactions scale with energy, the rate of graviton production is directly proportional to the energy scale, providing a unique probe into the extremely early universe via the stochastic GW background. Examples include GWs produced during the reheating epoch through $1\rightarrow3$ bremsstrahlung processes of massive particles~\cite{Nakayama:2018ptw,Huang:2019lgd,Ghoshal:2022kqp,Barman:2023ymn,Kanemura:2025rct}, $2\rightarrow2$ scattering, such as inflaton annihilation~\cite{Ema:2015dka,Ema:2016hlw} and scattering of thermal species~\cite{Ghiglieri:2015nfa,Ghiglieri:2020mhm,Ringwald:2020ist,Xu:2024fjl}. Additionally, light Schwarzschild or Kerr primordial black holes, due to their surface temperature being inversely proportional to their mass, can generate high-frequency GWs through Hawking evaporation~\cite{Anantua:2008am,Dolgov:2011cq,Dong:2015yjs,Ireland:2023avg,Choi:2024acs,Gross:2024wkl,Jiang:2025blz}.

In this work, motivated by the above discussions of dark matter, baryogenesis, and GW,
we propose a novel mechanism for the generation of GWs during cosmological first-order phase transitions: the braking of heavy particles as they traverse bubble walls. When a massive particle traverses a bubble wall during a first-order phase transition, it inevitably emits gravitons due to bremsstrahlung. This emission arises from the universal nature of gravitational interactions and is therefore unavoidable, regardless of the specific particle physics model.
Unlike conventional sources, this mechanism arises from the direct interaction between massive particles and the expanding bubble wall. We rigorously compute this gravitational radiation using quantum field theory. The resulting microscopic GW spectrum exhibits distinctive features that encode information about the particle mass spectrum and bubble wall dynamics: its peak frequency is tightly correlated with the bubble wall velocity, while its peak amplitude scales as the fourth power of the heavy particle mass. An interesting case arises when the particle mass is smaller than the temperature: the resulting GW spectrum exhibits a double-peaked structure. This additional peak frequency amplitude scales as the square of the mass.
These unique dependencies offer a new observational handle on particle physics beyond the Standard Model. We illustrate this mechanism within a specific model framework and demonstrate its viability. This emission arises from the universal nature of gravitational interactions and is therefore unavoidable, regardless of the specific particle physics model. Our findings enrich the landscape of phase transition GW sources and open new avenues for more directly probing heavy particle dynamics and new physics models in the early universe.

This paper is organized as follows. In Sec.~\ref{probabilitySec}, we analyze the probability for a scalar particle to emit a graviton via bremsstrahlung as it crosses the bubble wall during a first-order phase transition. In Sec.~\ref{GWEnergydensity}, we estimate the total energy density of the resulting GWs in the early universe. Sec.~\ref{GWspectrum} presents an analytic approximation for the GW spectrum today, obtained by reorganizing the phase space integration and considering the cosmological redshift. In Sec.~\ref{SecModel}, we illustrate the mechanism within a specific particle physics model and compute the corresponding GW signal. Sec.~\ref{Summary} contains our conclusions and a brief discussion of future directions. In Appendix~\ref{quantization}, we follow the method of Ref.~\cite{Azatov:2023xem} to quantize a real scalar field in the presence of a bubble wall, while in Appendix~\ref{probability}, we review the calculation of the bremsstrahlung probability as a massive particle traversing the bubble wall. \ccb{Finally, in Appendix C, we present an example of deriving the suppression region of the bremsstrahlung  process.}

\section{Graviton Bremsstrahlung Probability During Bubble Wall Crossing in First-Order Phase Transition}\label{probabilitySec}

We consider a strong first-order phase transition in the early Universe. As the temperature drops below the critical temperature \( T_c \) (more accurately, the nucleation temperature $T_n$), the false vacuum decays into the true vacuum through nucleation of bubbles. The potential energy difference between the true and false vacua drives the bubble walls to accelerate and expand. As the bubbles grow, plasma particles in the Universe cross the bubble walls, transitioning from the false vacuum into the true vacuum, where they acquire mass. From the perspective of the bubble wall frame, this mass acquisition implies a change in the particles' momentum, signaling a breakdown of momentum conservation. By analogy with classical physics, accelerated particles radiate GWs; microscopically, plasma particles crossing the bubble wall emit gravitons. In this work, we employ quantum field theory to rigorously compute the gravitational radiation produced by this process.

\begin{figure}
    \centering
    \includegraphics[width =0.66\linewidth]{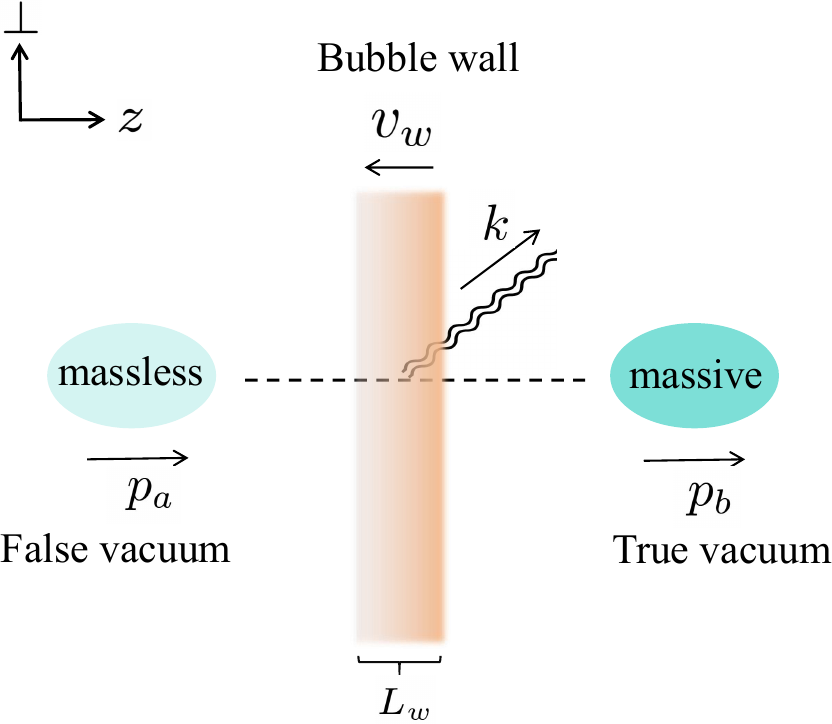}
    \caption{Schematic illustration of a scalar particle crossing the bubble wall during a first-order phase transition. As it enters the true vacuum region, the particle acquires mass and emits a graviton via bremsstrahlung. In the upper left corner of the figure, the selection of the coordinate axes is marked. In the wall frame, the direction of motion of the incident particle is taken to be along the positive $z$-axis. Relatively, from the perspective of the plasma frame, the bubble wall moves towards the particles along the negative $z$-axis. We denote by \( v_w \) and \( L_w \) the velocity and width of the bubble wall, respectively.
}
    \label{GWdemo}
\end{figure}

We focus on the process in which a scalar particle acquires a significant mass as it crosses the bubble wall and emits a graviton, denoted by \( s \rightarrow sg \). Our first task is to compute the probability that this process will occur. Since the probability is Lorentz invariant, we are free to perform the analysis in the rest frame of the bubble wall. A schematic illustration of this process is shown in Fig.~\ref{GWdemo}. In the highly relativistic regime, we assume that the incident scalar particle carries momentum purely in the longitudinal direction, perpendicular to the bubble wall. Under this assumption, we can formulate the kinematics of this \( 1 \rightarrow 2 \) process as the following
\begin{eqnarray}
    x &\equiv& \frac{E_k}{E_a},\\
    p_a &=& \left(E_a,0 ,0 ,\sqrt{E_a^2 - m^2(z)} \right),\\
    p_b &=& \Big((1-x)E_a,-k_{\bot},0,  p_b^z \Big),\quad (p_b^z)^2 = (1-x)^2 E^2_a - m^2(z) - k^2_\bot ,\\
    k &=& \Big(xE_a,k_\bot,0, k^z \Big),\quad (k^z)^2 =x^2 E^2_a  - k^2_\bot ,\\
    \varepsilon^\mu_\pm(k) &=& \frac{1}{\sqrt{2}}\left(0,\frac{k^z}{xE_a},\pm i, -\frac{k_\bot}{xE_a}  \right).
\end{eqnarray}
 Here, \( x \) denotes the energy fraction carried by the emitted graviton, defined as the ratio of the graviton energy \( E_k \) to the energy of the incoming scalar particle \( E_a \). The four-momentum of the initial scalar particle is denoted by \( p_a \), and that of the final scalar particle by \( p_b \). The mass of the heavy scalar particle is given by \( m(z) \), which depends on the spatial coordinate \( z \). In the broken phase, \( m(z) = m \), while in the symmetric phase, we assume \( m(z) = 0 \). The four-momentum of the emitted graviton is denoted by \( k \), and \( \varepsilon^\mu_\pm(k) \) represents its two physical polarization vectors. For convenience, we assume that the plasma is moving along the positive \( z \)-axis, and exploit the symmetry between the two transverse components to simplify the final-state particle's four-momentum expression.

In Ref.~\cite{Azatov:2023xem}, a general framework for quantizing scalar and vector fields in the presence of a bubble wall background was developed, and applied to the computation of NLO friction acting on the wall. Within that context, the probability for a scalar particle to radiate a vector boson while crossing the bubble wall was well-defined. Analogously, a scalar particle may also emit a graviton during this process, and the probability of such a process can be computed in a manner similar to the emission of a massless gauge boson. Consequently, in the presence of a bubble wall, the probability of the process \( s \rightarrow sg \) is given by
\begin{eqnarray}
   \int \mathrm{d}P_{s\rightarrow sg} 
    &=& \int \frac{\mathrm{d}^3 p_b}{(2\pi)^3 2E_{b}} \int \frac{\mathrm{d}^3 k}{(2\pi)^32E_{k}}  \frac{1}{2E_a} (2\pi)^3 \delta^{(2)}\left(\sum \vec{p}_{\bot} \right)  \delta\left(\sum E \right)\left(|\mathcal{M}_R|^2 + |\mathcal{M}_L|^2\right)\nonumber\\
    &=& \int \frac{\mathrm{d}^3 k}{(2\pi)^3}\frac{1}{2p_{b,s}^z}\frac{x}{(2E_k)^2}\left( |\mathcal{M}_R|^2 +|\mathcal{M}_L|^2  \right).
\end{eqnarray}
It is important to note that within the framework of Ref.~\cite{Azatov:2023xem}, $\mathrm{d}^3p_b = \mathrm{d}p^z_{b,s}\mathrm{d}^2p_{b,\bot}$, where $p^z_{b,s} = \sqrt{(1-x)^2E^2_a - k^2_\bot}$ is the longitudinal momentum value of the final-state scalar particle in the symmetric phase (i.e., at $z=-\infty$). The integration range for $p^z_{b,s}$ is taken to be \( [0, \infty) \). In this formalism, transmission (right-moving) and reflection (left-moving) are treated as distinct final states, with corresponding matrix elements denoted by \( \mathcal{M}_R \) and \( \mathcal{M}_L \), respectively. In our coordinate system, the initial state is a right-moving wave, and thus the matrix elements \( \mathcal{M}_{R,L} \) can be expressed as     
\begin{eqnarray}
 \mathcal{M}_I &=& \int_{-\infty}^{\infty} \mathrm{d} z~ \chi_{R}(z,p_{a,s}^z)\zeta_{I}^*(z, p_{b,s}^z)\chi^*(z,k^z) V(z)  ,\quad I = R,L.
 \end{eqnarray}
 Here, \( V(z) \) denotes the amplitude of the process at position \( z \), $p^z_{a,s} = E_a$ is the longitudinal momentum value of the initial-state scalar particle in the symmetric phase. \( \chi_{R}(z, p_{a,s}^z) \), \( \zeta_{I}(z, p_{b,s}^z) = \sum_J \chi^*_J(z, p_{b,s}^z)\epsilon^{IJ} \), and \( \chi(z, k^z) = e^{ik^z z} \) represent the mode functions of the incoming particle, the outgoing particle and the graviton respectively. The antisymmetric tensor \( \epsilon^{IJ} \) is defined such that \( \epsilon^{IJ} = 0 \) for \( I = J \) and \( \epsilon^{IJ} = 1 \) for \( I \neq J \). Considering that the final-state scalar particle remains highly relativistic after acquiring mass, the reflection amplitude is suppressed in this regime, satisfying \( \mathcal{M}_L \ll \mathcal{M}_R \). Consequently, it is appropriate to employ the Wenzel–Kramers–Brillouin (WKB) approximation for evaluating the matrix element. In this work, we neglect the contribution from reflection and compute the matrix element using the WKB method. Consequently, we omit the subscripts \( R, L \), and the matrix element is simplified to
 \begin{eqnarray}
     \mathcal{M}_L &\simeq& 0,\\
     \mathcal{M}_R &\simeq& \mathcal{M}^{\rm WKB} = \int_{-\infty}^{\infty} \mathrm{d} z~ \chi(z,p_{a,s}^z)\chi^*(z, p_{b,s}^z)\chi^*(z,k^z) V(z).\label{9shi}
 \end{eqnarray}
 
 The mode function $\chi(z,p_s^z)$, under the WKB approximation, can be written as
 \begin{eqnarray}
     \chi(z,p_s^z) \simeq \sqrt{\frac{p^z_s}{p^z(z)}}\mathrm{exp}\left[i\int_0^z \mathrm{d}z^\prime~p^z(z^\prime) \right].
 \end{eqnarray}
In the ultra-relativistic limit, the prefactor can be approximated as unity. Under the WKB approximation, the interaction vertex between the scalar particle and the graviton retains the same form as in the absence of a bubble wall, with relevant vertex structures provided in Ref.~\cite{Bjerrum-Bohr:2004qcf}. However, the momentum becomes $z$-dependent due to the presence of the bubble wall, and the amplitude of the splitting process occurring at position $z$ is consequently modified as
 \begin{eqnarray}
  V_{\pm}(z) &=& -\frac{1}{2} i \kappa  m^2(z) \varepsilon^2_\pm (k)+\frac{1}{2} i \kappa  \varepsilon_\pm^2 (k) (p_a\cdot p_b  )-i \kappa  (p_a \cdot  \varepsilon_\pm (k)) (p_b \cdot  \varepsilon_\pm (k))\nonumber\\
   &\simeq&  
  -\frac{i\kappa k^2_\bot}{2x^2E_a}\big[p_a^z(z) -  \Delta p^z(z)\big], \label{vertexfunc}
\end{eqnarray}
where the gravitational interaction coupling $\kappa^2 = 32\pi/M^2_{\rm pl}$ with the Planck mass $M_{\rm pl} = 1.22\times 10^{19}~\mathrm{GeV}$, and the longitudinal momentum transfer $ \Delta p^z(z) = p_a^z(z) - p_b^z(z) - k^z$. Here we have used the ultra-relativistic approximation $E_a \gg m$.

{If the longitudinal momentum transfer is written as $\Delta p^z(z) = \Delta p^z(+\infty) + \Delta_z$, Eq.~\eqref{9shi} becomes
\begin{align}
    \mathcal{M}^{\rm WKB}= \int_{-\infty}^{\infty}\mathrm{d}z~V(z)e^{i\int_0^z\mathrm{d}z^\prime\Delta_z} e^{i\Delta p^z(+\infty)z},
\end{align}
which is a Fourier transform formula. Then we assume that $m(z)$ only varies inside the wall (i.e., $z\in [0,L_w]$, where $L_w$ denotes the width of the bubble wall), while rapidly reaching an asymptotic constant value outside the wall. According to the basic properties of the Fourier transform, therefore, when $\Delta p^z(+\infty) L_w \gg 1$, $\mathcal{M}^{\rm WKB}\to 0$. This condition is also referred to as the non-adiabaticity condition~\cite{Azatov:2024crd}, and it
is also intuitively understandable physically: large momentum transfer means the transition proceeds over a distance scaling $\sim 1/\Delta p^z(+\infty)$. For distances $\ll L_w$, translational symmetry along the $z$-direction is restored, thereby prohibiting the transition radiation (see Appendix~\ref{non-adiabaticity} for a check by using a wall ansatzes). Consequently, independent of the specific shape of the wall, the dominant contribution region of the transition radiation amplitude is $\Delta p^z(+\infty) L_w \lesssim 1$.
On the other hand, we can separate the contributions of the amplitude inside and outside the wall, given by
\begin{align}
\mathcal{M}^{\rm WKB} =&
\int_{-\infty}^0 \mathrm{d}z~V_s e^{i \Delta p^z_s z} 
+ e^{i \int_0^{L_w} \mathrm{d}z^\prime\Delta p(z^\prime)} \int_0^\infty \mathrm{d}z~V_b e^{i \Delta p^z_b z}+ \int_0^{L_w} \mathrm{d}z~V(z) e^{i \int_0^z \mathrm{d}z^\prime\Delta p(z^\prime)}\nonumber\\
\simeq& \int_{-\infty}^0 \mathrm{d}z~V_s e^{i \Delta p^z_s z} 
+  \int_0^\infty \mathrm{d}z~V_b e^{i \Delta p^z_b z}+ \int_0^{L_w} \mathrm{d}z~V(z) e^{i \int_0^z \mathrm{d}z^\prime\Delta p(z^\prime)},\label{eq:MWKB}
\end{align}
where $V_s \equiv V(-\infty)$ ($V_b \equiv V(+\infty)$), and $\Delta p_s^z \equiv \Delta p^z(-\infty)$ ($\Delta p^z_b \equiv \Delta p^z(+\infty)$) to simplify notation. Under the condition $\Delta p^z_b L_w \lesssim 1$, the last term in Eq.~\eqref{eq:MWKB} can be neglected compared to the first two terms. Thus we reproduce the Bodeker–Moore formula~\cite{Bodeker:2017cim}
\begin{eqnarray}
    \mathcal{M}^{\rm WKB} &\simeq& \frac{V_s}{i\Delta p^z_s} - \frac{V_b}{i\Delta p^z_b} 
    \simeq -\frac{\kappa k_\bot^2}{2x^2} \left( \frac{\Delta p^z_b - \Delta p^z_s}{\Delta p^z_s\Delta p^z_b} \right). \label{BM}
\end{eqnarray}

}
Under the soft-collinear approximation, defined by $k_\bot \ll xE_a$, $x \ll 1$, and $k^z > 0$, the leading-order expressions are
\begin{eqnarray}
    \Delta p^z_b - \Delta p^z_s &\simeq& \frac{x m^2}{2E_a},\quad
    \Delta p^z_b \Delta p^z_s \simeq \frac{k_\bot^2 (k_\bot^2 + x^2 m^2)}{4x^2 E_a^2}.
\end{eqnarray}
Taking the modulus squared of the amplitude and summing over polarization states, we obtain the radiation probability
\begin{eqnarray}
    \int \mathrm{d}P_{s \rightarrow sg} 
    &\simeq& \int \frac{\mathrm{d}^3k}{(2\pi)^3 2E_k}~ \mathcal{P}(k) \, 
    \Theta\left(p_{b,s}^z - L_w^{-1}\right) 
    \Theta\left(L_w^{-1} - \ccb{\Delta p^z_b} \right) 
    \Theta\left(k^z\right), \label{probabilityEq} \\
    \mathcal{P}(k) &\equiv& \frac{\kappa^2 m^4 E_a^2 E_k^2}{2(E_a^2 k_\bot^2 + m^2 E_k^2)^2},
\end{eqnarray}
The first two Heaviside functions in Eq.~\eqref{probabilityEq} enforce the WKB condition $p_{b,s}^z  > L_w^{-1}$ and the non-adiabaticity condition \ccb{$\Delta p^z_bL_w  < 1$.} We can see that the radiation probability is maximized as \(k_{\bot} \to 0\), which implies that the dominant radiation as the particle traverses the bubble wall consists of collinear gravitons.

\section{Gravitational wave energy density}\label{GWEnergydensity}

With the radiation probability in hand, we can estimate the total radiated energy density. It is important to note that GWs are observed in the plasma frame. The energy and momentum of the graviton in this frame can be obtained via a Lorentz transformation. Throughout the following discussion, we denote quantities in the plasma frame using variables with a tilde,
\begin{eqnarray}
    \tilde{E}_k = \gamma(E_k - v_w
    k^z),\quad \tilde{k}^z = \gamma(k^z - v_w E_k),\quad \tilde{k}_\bot = k_\bot,
\end{eqnarray}
Here, $v_w$ denotes the bubble wall velocity in the plasma frame, with the corresponding Lorentz factor given by $\gamma \equiv \sqrt{1/(1 - v_w^2)}$. The total GW energy density is computed as the product of the particle number density and the expectation value of the graviton energy, expressed as
\begin{eqnarray}
    \rho_{\rm GW} = \int \mathrm{d}^3\tilde{p}_a~f_a(\tilde{p}_a)\braket{\tilde{E}_k},\label{GWenegy}
\end{eqnarray}
where \(
    f_a(\tilde{p}_a) = 1/\left(e^{\tilde{p}_a/T} - 1\right)
\)
is the distribution function of the thermal plasma at temperature $T$.

Due to the Lorentz invariance of the probability, we can boost to the wall frame to compute the expectation value of the graviton energy, given by $\braket{\tilde{E}_k} = \braket{\gamma(E_k - v_w k^z)}$.
To compute the expectation value of the graviton energy, we simplify the phase space integration region as follows. The phase space integration can be expressed as $\int\mathrm{d}^3k = \pi\int_0^{E_k^{\rm max}} \mathrm{d}E_k\int_0^{K_{\bot}^2} \mathrm{d}k^2_\bot$. Here, we recast the Heaviside function into \(E_k^{\rm max}\) and \(K_\bot\), the upper limits of the integrals over \(E_k\) and \(k_\bot\).
From the WKB condition $p_{b,s}^z > L_w^{-1}$, we obtain the constraint $k^2_\bot < (E_a - E_k)^2 - L_w^{-2}$. Under the approximations $x \ll 1$ and $E_a \gg m \sim L_w^{-1}$, we have $E_k^2 < (E_a - E_k)^2 - L_w^{-2} - m^2$. Consequently, the condition $p_{b,s}^z > L_w^{-1}$ is always satisfied and the upper bound of the $E_k$ integration is given by
\begin{eqnarray}
    E_k^{\rm max} = \frac{1}{2}\left(E_a - \frac{L_w^{-2} + m^2}{E_a} \right) \sim \frac{1}{2} E_a.
\end{eqnarray}
Additionally, using the non-adiabaticity condition  $\ccb{\Delta p^z_b} \simeq E_k - \sqrt{E_k^2 - k^2_\bot} < L_w^{-1}$, we derive $k^2_\bot < L_w^{-1}(2E_k - L_w^{-1})$. In summary, these conditions define the allowed phase space region for the integration:
\begin{equation}
    \begin{aligned}[b]
            k_\bot^2 &< L_w^{-1}(2E_k - L_w^{-1})&,& \quad \mathrm{if}\quad L_w^{-1} < E_k < E_k^{\rm max},\\
    k_\bot^2 &<  E_k^2 &,& \quad \mathrm{if} \quad 0<E_k< L_w^{-1} .
    \end{aligned}\label{phasespace}
\end{equation}
Accordingly, the average energy of the graviton is given by 
\begin{eqnarray}
     \braket{\tilde{E}_k} &=& \braket{\gamma(E_k - v_w k^z)}\nonumber\\
    &=& \frac{\kappa^2 m^4\gamma}{32 \pi^2} \left[\int_0^{L_w^{-1}} \mathrm{d}E_k\int_0^{E_k^2} \mathrm{d}k^2_\bot~\left(\frac{ E_a^2E_k^3}{k^z(E_a^2 k^2_\bot + m^2 E_k^2)^2} - v_w \frac{ E_a^2E_k^2 }{(E_a^2 k^2_\bot + m^2 E_k^2)^2}\right)\right.\nonumber\\
    &&+ \left. \int_{L_w^{-1}}^{E_k^{\rm max}} \mathrm{d}E_k\int_0^{L_w^{-1}(2E_k - L_w^{-1})} \mathrm{d}k^2_\bot ~\left(\frac{ E_a^2E_k^3}{k^z(E_a^2 k^2_\bot + m^2 E_k^2)^2}    -   v_w\frac{ E_a^2E_k^2 }{(E_a^2 k^2_\bot + m^2 E_k^2)^2}\right)          \right]\nonumber\\
    &\simeq&\frac{\kappa^2 m^4}{32 \pi^2}\left( \frac{E_a}{4\gamma m^2} + \frac{\gamma}{4E_a}\ln\frac{8E_aL_w^{-1}}{m^2} \right).\label{Ektaverage}
\end{eqnarray}

The integration over the plasma distribution function can also be boosted to the bubble wall frame for convenience. Notably, the initial energy variable $E_a$ appearing in the previous equations corresponds to the longitudinal momentum component $p_{a,s}^z$ in the plasma distribution function. Taking this into account, the total radiated energy density becomes
\begin{eqnarray}
    \rho_{\rm GW} &=& \int \frac{\mathrm{d}p^z_{a,s}}{\gamma 2\pi}\int \frac{\mathrm{d}^2p_{a,\bot}}{(2\pi)^2}  ~ \frac{\braket{\tilde{E}_k}}{e^{\gamma\left(\sqrt{(p_{a,s}^z)^2 + p_{a,\bot}^2}- v_w p_{a,s}^z\right)/T}-1}\nonumber\\
     &\simeq& -\int_M^\infty \frac{\mathrm{d}p^z_{a,s}}{ 8\pi^2} ~\frac{2 T p_{a,s}^z}{\gamma }\ln \left(1-e^{-\frac{p_{a,s}^z}{2 \gamma  T}}\right)\frac{\kappa^2 m^4}{32 \pi^2}\left( \frac{p^z_{a,s}}{4\gamma^2m^2} + \frac{1}{4p^z_{a,s}}\ln\frac{8p^z_{a,s}L_w^{-1}}{m^2} \right)\nonumber\\
     &\simeq& \frac{\kappa^2m^4 T^2}{128\pi^4}\left[\frac{2\pi^4 T^2}{45m^2} 
 + \ln \frac{2\pi^2 M L_w^{-1}}{3m^2}   + \sum_{n=1}^\infty\frac{1}{2n^2} \Gamma\left(0,\frac{nM}{2\gamma T}\right)  \right].
\end{eqnarray}
The lower bound of the longitudinal momentum component $p_{a,s}^z$ is given by 
\ccb{
\begin{equation}
M =  \sqrt{m^2 + 2L_w^{-2}} + L_w^{-1},
\end{equation}}which follows from the condition $E_k^{\rm max} > L_w^{-1}$. The function $\Gamma(0,x)$ denotes the incomplete Gamma function, which admits the approximation $\Gamma(0,x) \sim \ln(1/x)$ for $x \ll 1$. To obtain an analytically tractable expression, we perform a Taylor expansion of the logarithmic term $\ln \left(1 - e^{-p_{a,s}^z / (2 \gamma T)}\right)$.

Under the assumption that $m \gg T$, the leading-order contribution to the GW energy density is found to be 
\begin{equation}
\rho_{\rm GW} \sim \frac{\kappa^2 m^4 T^2}{256\pi^4} \ln\left( \frac{2\gamma T}{M} \right). 
\end{equation}

\section{Gravitational wave spectra}\label{GWspectrum}

The number density of radiated gravitons can be obtained straightforwardly by multiplying the plasma particle number density with the graviton emission probability. Retaining the phase space integral over graviton momenta allows us to derive the corresponding graviton distribution function. Since our goal is to obtain the graviton distribution function in the plasma frame, it is necessary to express the emission probability given in Eq.~\eqref{probabilityEq} using variables defined in the plasma frame. The Lorentz invariance of the probability implies that the radiation probability in the plasma frame can be obtained through a straightforward change of variables. Therefore, we have
\begin{eqnarray}
 \int \mathrm{d}P_{s\rightarrow sg} &=&   \int\frac{\mathrm{d}^3k}{(2\pi)^32E_k}~\mathcal{P}(k) = \int \frac{\mathrm{d}^3\tilde{k}}{(2\pi)^3 2\tilde{E}_k}~\tilde{\mathcal{P}}(\tilde{k}) \nonumber\\
&=& \frac{\kappa^2 m^4 \gamma^2}{16\pi^2E_a^2}\int\mathrm{d}\tilde{E}_k\int_{k_{\rm min}}^{k_{\rm max}} \mathrm{d}\tilde{k}^z~\frac{(\tilde{E}_k + v_w\tilde{k}^z)^2}{\big[\tilde{E}_k^2 - (\tilde{k}^z)^2 + \frac{m^2\gamma^2}{E_a^2}(\tilde{E}_k + v_w\tilde{k}^z)^2\big]^2}.\quad \label{gailvbiaoasj}
\end{eqnarray}
Since we are primarily interested in the energy of the graviton, we perform the integration over the longitudinal momentum component $\tilde{k}^z$, retaining only the energy variable $\tilde{E}_k$. The integration limits, denoted by $k_{\rm min}$ and $k_{\rm max}$, are determined by the phase space region in the bubble wall frame as defined in Eq.~\eqref{phasespace}, with the additional condition that $k^z > 0$. The overall phase space can be partitioned into six distinct regions, as summarized in Table~\ref{phase-spacePlasmaFrame}.

\begin{table}[]
    \centering
    \renewcommand\arraystretch{2}
    \begin{tabular}{|c|c|c|}
\hline 
 & $p_a^z$-$\tilde{E}_k$ plane & Range of $\tilde{k}^z$ \\ 
\hline \rule{0pt}{30pt}
   A  & $\frac{E_k^{\rm max}}{(1+v_w)\gamma} < \gamma L_w^{-1}$, $0<\tilde{E}_k < \frac{E_k^{\rm max}}{(1+v_w)\gamma} $ & $\begin{aligned}
       k_{\rm min} &= -v_w \tilde{E}_k \approx -\tilde{E}_k,\\
       k_{\rm max} &= \tilde{E}_k
   \end{aligned} $ \\ 
   \hline \rule{0pt}{30pt}
   B & $\frac{E_k^{\rm max}}{(1+v_w)\gamma} < \gamma L_w^{-1}$, $\frac{E_k^{\rm max}}{(1+v_w)\gamma}<\tilde{E}_k< \gamma L_w^{-1}$ & $\begin{aligned}
       k_{\rm min} &= -v_w \tilde{E}_k \approx -\tilde{E}_k, \\
       k_{\rm max} &= \frac{1}{v_w}\left(\frac{E_k^{\rm max}}{\gamma} -\tilde{E}_k \right)\approx \frac{E_k^{\rm max}}{\gamma} -\tilde{E}_k 
   \end{aligned}$ \\[0.5cm]
   \hline
    \rule{0pt}{30pt} C & $\frac{E_k^{\rm max}}{(1+v_w)\gamma} < \gamma L_w^{-1}$, $\gamma L_w^{-1}< \tilde{E}_k < \frac{E_k^{\rm max}}{(1+v_w)\gamma} + v_w \gamma L_w^{-1}$ & $\begin{aligned}
        k_{\rm min} &= \tilde{E}_k - (1+v_w)\gamma L_w^{-1} \approx \tilde{E}_k - 2\gamma L_w^{-1}, \\
          k_{\rm max} &= \frac{1}{v_w}\left(\frac{E_k^{\rm max}}{\gamma} -\tilde{E}_k \right)\approx \frac{E_k^{\rm max}}{\gamma} -\tilde{E}_k
   \end{aligned}$ \\[0.5cm]
   \hline 
  \rule{0pt}{30pt} D & $\frac{E_k^{\rm max}}{(1+v_w)\gamma} > \gamma L_w^{-1}$, $0<\tilde{E}_k < \gamma L_w^{-1}$ & $\begin{aligned}
       k_{\rm min} &= -v_w \tilde{E}_k \approx -\tilde{E}_k ,\\
          k_{\rm max} &= \tilde{E}_k
   \end{aligned}$ \\
   \hline 
  \rule{0pt}{30pt}  E & $\frac{E_k^{\rm max}}{(1+v_w)\gamma} > \gamma L_w^{-1}$, $\gamma L_w^{-1} <\tilde{E}_k < \frac{E_k^{\rm max}}{(1+v_w)\gamma}$ & $\begin{aligned}
       k_{\rm min} &= \tilde{E}_k - (1+v_w)\gamma L_w^{-1}\approx \tilde{E}_k -2\gamma L_w^{-1},\\
           k_{\rm max} &= \tilde{E}_k
   \end{aligned}$ \\
   \hline 
   \rule{0pt}{30pt} F & $\frac{E_k^{\rm max}}{(1+v_w)\gamma} > \gamma L_w^{-1}$, $\frac{E_k^{\rm max}}{(1+v_w)\gamma} < \tilde{E}_k < \frac{E_k^{\rm max}}{(1+v_w)\gamma} + v_w \gamma L_w^{-1}$ & $\begin{aligned}
       k_{\rm min} &= \tilde{E}_k - (1+v_w)\gamma L_w^{-1}\approx \tilde{E}_k -2\gamma L_w^{-1},\\
           k_{\rm max} &= \frac{1}{v_w}\left(\frac{E_k^{\rm max}}{\gamma} - \tilde{E}_k \right) \approx \frac{E_k^{\rm max}}{\gamma} - \tilde{E}_k
   \end{aligned}$\\[0.5cm]
   \hline
\end{tabular}
    \caption{Phase space integration regions. In the leftmost column of the table, we label the integration regions using the identifiers A, B, C, D, E, and F.}
    \label{phase-spacePlasmaFrame}
\end{table}

For each region, we first perform the integration over the longitudinal component $\tilde{k}^z$, simplifying the expression. As a result, the graviton emission probability takes the form
\begin{eqnarray}
\int \mathrm{d}P_{s\rightarrow sg} &\approx& \frac{\kappa^2 m^4 }{16\pi^2E_a^2}\sum_{i}\int_{i}\mathrm{d}\tilde{E}_k~\tilde{P}_{i}(\tilde{E}_k) ,\quad i={\rm A,B,C,D,E,F},
\end{eqnarray}
with
\begin{eqnarray}
    \tilde{P}_{\rm A}(\tilde{E}_k) =\tilde{P}_{\rm D}(\tilde{E}_k) &\approx&  \frac{E_a^2}{2m^2 \tilde{E}_k},\\
    \tilde{P}_{\rm B}(\tilde{E}_k) &\approx& \frac{\gamma^2 E_a^2 E_k^{\rm max}}{2 \tilde{E}_k \left((2\gamma\tilde{E}_k -E_k^{\rm max})E_a^2 +\gamma^2 m^2 E_k^{\rm max}\right)},\\
     \tilde{P}_{\rm C}(\tilde{E}_k) = \tilde{P}_{\rm F}(\tilde{E}_k) &\approx& \frac{\gamma E_a^4 \left(E_k^{\rm max} + 2\gamma^2 L_w^{-1} - 2\gamma \tilde{E}_k\right)}{2\left(\gamma^2 m^2 E_k^{\rm max} + (2\gamma \tilde{E}_k - E_k^{\rm max} )E_a^2 \right) \left(L_w^{-1}E_a^2 +\gamma m^2(\tilde{E}_k - \gamma L_w^{-1})\right)},\quad\,\\
      \tilde{P}_{\rm E}(\tilde{E}_k) &\approx&  \frac{E_a^4 L_w^{-1}}{2m^2 \tilde{E}_k\left(E_a^2 L_w^{-1} + \gamma m^2 (\tilde{E}_k - \gamma L_w^{-1})   \right)}.
\end{eqnarray}

Next, we retain the integration over the graviton energy variable $\tilde{E}_k$ while performing the integral over the plasma distribution function. As before, it is convenient to boost the calculation to the bubble wall frame. In this frame, we first integrate over the transverse momentum component $p_{a,\bot}$, which is independent of the phase space regions discussed above, yielding
\begin{eqnarray}
    \rho_{\rm GW} &=&\frac{\kappa^2 m^4 T}{64\pi^4 } \sum_i \iint_{i}\mathrm{d}\tilde{E}_k\mathrm{d}p_{a}^z~\left[-\frac{\tilde{P}_i(\tilde{E}_k )\tilde{E}_k}{\gamma^2 p_{a}^z}\ln\left(1-e^{-\frac{p_{a}^z}{2\gamma T}} \right) \right].
\end{eqnarray}
To simplify the notation, we omit the subscript ``$s$'' from the longitudinal momentum component \(p^z_{a,s}\) in the integral of the plasma distribution function.

\begin{figure}
    \centering
    \includegraphics[width=0.8\linewidth]{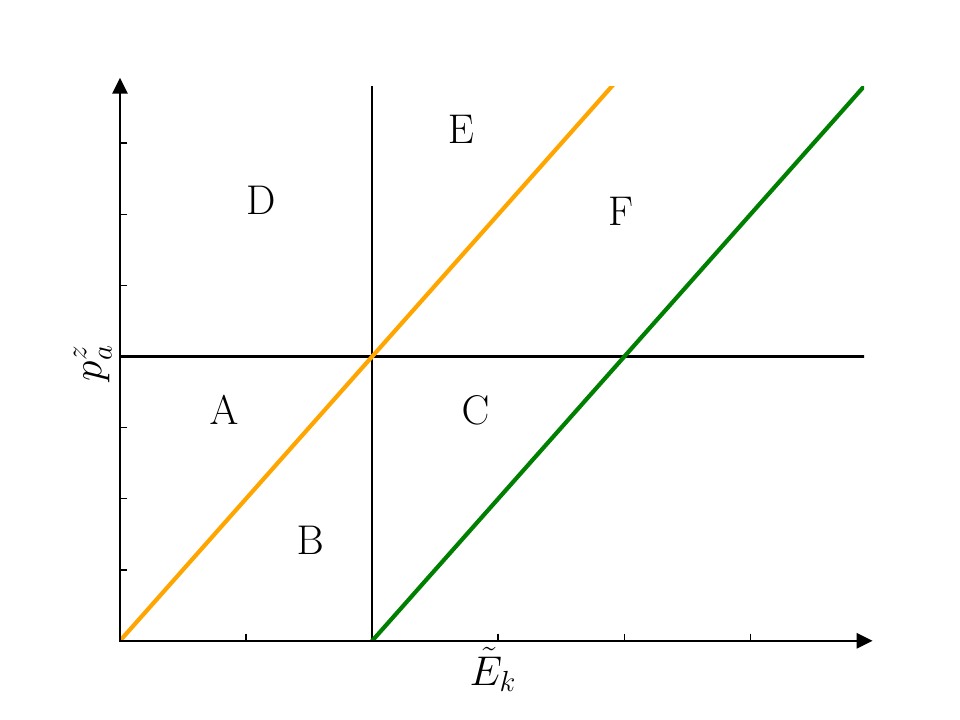}
    \caption{Integration region in the $p_a^z$–$\tilde{E}_k$ plane. The orange line corresponds to $p_a^z = 4\gamma \tilde{E}_k$, while the green line denotes $p_a^z = 4\gamma \tilde{E}_k - 4\gamma^2 L_w^{-1}$. The six distinct regions share a common boundary point at $(\gamma L_w^{-1},\, 4\gamma^2 L_w^{-1})$.}
    \label{paz-Ekt}
\end{figure}

Before integrating over $p_{a}^z$, we first discuss how to simplify the integration region. To streamline the calculation, we adopt the approximation $E_k^{\rm max} \approx \frac{1}{2}p_{a}^z$. Under this assumption, the integration region in the $p_{a}^z$–$\tilde{E}_k$ plane is depicted in Fig.~\ref{paz-Ekt}.
In this figure, the horizontal axis represents the graviton energy $\tilde{E}_k$, while the vertical axis corresponds to the longitudinal momentum component $p_a^z$. The orange line marks the relation $p_a^z = 4\gamma \tilde{E}_k$, and the green line corresponds to $p_a^z = 4\gamma \tilde{E}_k - 4\gamma^2 L_w^{-1}$. The six phase-space regions intersect at a common boundary point given by $(\gamma L_w^{-1},\, 4\gamma^2 L_w^{-1})$. The lower boundary of the integration region is nominally set by the previously discussed limit $p_a^z > M$. However, it will impose the condition $\tilde{E}_k > M/(4\gamma)$, since we use the approximation $E_k^{\rm max} \approx \frac{1}{2}p_a^z$. On the other hand, since the dominant contribution to the integrand arises around $p_a^z \sim \gamma T$, the impact of neglecting small $p_a^z$ values is expected to be minimal. Consequently, the lower limit of $p_a^z$ can be safely approximated as zero without significantly affecting the resulting power spectrum. To verify this, we temporarily retain the lower bound of $p_a^z$ as $M$, and then take the limit $M \rightarrow 0$ in the final integration result.
It is also worth noting that the integrands in regions A and D are identical, allowing these to be merged, with a similar merging applicable for regions C and F.

Based on the discussion of the integration regions above, the GW spectrum can be computed by dividing the phase space into two distinct regimes: the relatively low-frequency region (A, B, D) and the relatively high-frequency region (C, E, F). We emphasize once again that the energy variable $E_a$ appearing in the emission probability corresponds to the longitudinal momentum component $p_a^z$ used in the distribution function.

Thus, we can obtain the GW energy density at production time,
\begin{eqnarray}
    \rho_{\rm GW} = \frac{\kappa^2 m^4 T}{64\pi^4 } \left[\int_{\rm low}\mathrm{d}\tilde{E}_k~I_{\rm low}(\tilde{E}_k)    + \int_{\rm high}\mathrm{d}\tilde{E}_k~I_{\rm high}(\tilde{E}_k)              \right],
\end{eqnarray}
with the relatively low-frequency part
\begin{eqnarray}
    I_{\rm low}(\tilde{E}_k) &=& \sum_{n=1}^{\infty} \frac{ \xi e^{n\xi}\Big[  \mathrm{Ei}\left(-n\epsilon -n\xi   \right) -  \mathrm{Ei}\left(-n\eta \right) \Big] + \eta e^{-n\eta}  \Big[\mathrm{Ei}\left(-n\epsilon + n\eta \right)  -  \mathrm{Ei}\left(n\xi \right) \Big] }{4n (\xi + \eta)} \nonumber\\
    &&+ \frac{2T^2}{m^2}\Big[(\eta -\xi)\mathrm{Li}_2\left(e^{\xi - \eta} \right) + \mathrm{Li}_3\left(e^{\xi - \eta} \right) \Big], \label{lowf}\\
    \xi &=&\frac{\sqrt{4\tilde{E}_k^2 + m^2} - 2\tilde{E}_k}{2T},\quad \eta = \frac{\sqrt{4\tilde{E}_k^2 + m^2} + 2\tilde{E}_k}{2T},\quad \epsilon = \frac{M}{2\gamma T},
\end{eqnarray}
where $\mathrm{Ei}(x)$ denotes the exponential integral function and $\mathrm{Li}_n(x)$ represents the polylogarithm function. To streamline the analytical expression, we introduce three dimensionless parameters, $\xi$, $\eta$, and $\epsilon$. The second term in Eq.~\eqref{lowf} arises from contributions in regions A and D, and it becomes exponentially suppressed when $\tilde{E}_k \gg T$. Notably, in the limit $\epsilon \rightarrow 0$, the function $I_{\rm low}(\tilde{E}_k)$ exhibits no significant variation, as mentioned previously. In the relatively high-frequency regime, the contribution resembles that of regions A and D, exhibiting significant impact only within the domain $4 \tilde{E}_k - 4\gamma L_w^{-1} < T$. Consequently, the high-frequency gravitons with $\tilde{E}_k > \gamma L_w^{-1}$ are heavily suppressed, effectively truncated from the spectrum. Thus, we neglect them in the subsequent calculations. When $m\gg T$, upon integrating over $\tilde{E}_k$, the dominant contribution arises from the first term in Eq.~\eqref{lowf}. By substituting the variable $\tilde{E}_k \rightarrow xT$, we recover the earlier result $\rho_{\rm GW} \propto \kappa^2 m^4 T^2$. Furthermore, the leading-order behavior of the remaining integral is expected to scale proportionally with $\ln \gamma$.

Taking into account cosmological redshift effects, the present-day energy density and frequency of the GW can be expressed accordingly as
\begin{eqnarray}
    \rho_{\rm GW,0} &=& \left(\frac{a_*}{a_0}\right)^4 \rho_{\rm GW},\\
    f_0 &=& \frac{a_*}{a_0}\frac{\tilde{E}_k}{2\pi},
\end{eqnarray}
where the cosmic scale factor today is defined as $a_0 = 1$. The corresponding scale factor at the production time of GW can be estimated by the entropy conservation as 
\begin{equation}
a_*\simeq \left(\frac{g_{{\rm eq}, s}T_{\rm eq}^3}{g_{*,s} T^3}\right)^{1/3}a_{\rm eq} .
\end{equation}
At the epoch when matter and radiation densities are equal, the entropy degrees of freedom are approximately given by $g_{{\rm eq},s} \simeq 3.94$~\cite{Husdal:2016haj}, with a corresponding temperature of $T_{\rm eq} \simeq 0.80~\mathrm{eV}$ and scale factor $a_{\rm eq} \simeq 2.9 \times 10^{-4}$~\cite{ParticleDataGroup:2024cfk}. In our analysis, we approximate the entropy degrees of freedom by the relativistic degrees of freedom in the matter sector, adopting $g_{*s} \simeq g_* = 106.75$ at temperature $T$. Finally, we obtain the GW spectrum for this new GW source,
\begin{eqnarray}
    h^2\Omega_{\rm GW,0}^{\rm brakes}(f_0) &=& \frac{h^2}{\rho_{c,0}}\frac{\mathrm{d} \rho_{\rm GW,0}}{\mathrm{d}\ln f_0} =\frac{h^2}{\rho_{c,0}}\frac{\mathrm{d} \rho_{\rm GW,0}}{\mathrm{d}\ln \tilde{E}_k}  \nonumber\\
    &\simeq& \left(\frac{a_*}{a_0}\right)^4\frac{h^2}{\rho_{c,0}}\frac{ \kappa^2m^4 T}{64 \pi^4}~\tilde{E}_k ~I_{\rm low}\left(\tilde{E}_k\right) \Theta\left(\gamma L_w^{-1} - \tilde{E}_k\right)
    \nonumber\\
    &=& 6.91 \times 10^{-20}~ \left(\frac{3.94}{g_{*,s}}\right)~\left(\frac{m}{T}\right)^2~\left(\frac{m}{10^{13}~\mathrm{GeV}}\right)^2~\left(\frac{f_0}{10^{10}~\mathrm{Hz}}\right)\nonumber\\
    &&\times I_{\rm low}\left(\tilde{E}_k\right)~\Theta\left(\gamma L_w^{-1} - \tilde{E}_k\right),\\
    \tilde{E}_k &=& 2.71\times 10^{24}~\left(\frac{g_{*,s}}{3.94}\right)^{1/3}~\left(\frac{T}{10^{11}~\mathrm{GeV}}\right)~f_0,
\end{eqnarray}
with the critical energy density today $\rho_{c,0} = 8.07\times10^{-47} h^2~\mathrm{GeV}^4$ and dimensionless Hubble parameter $h=0.674$~\cite{Planck:2018vyg}.

\begin{figure}
    \centering
    \includegraphics[width=0.88\linewidth]{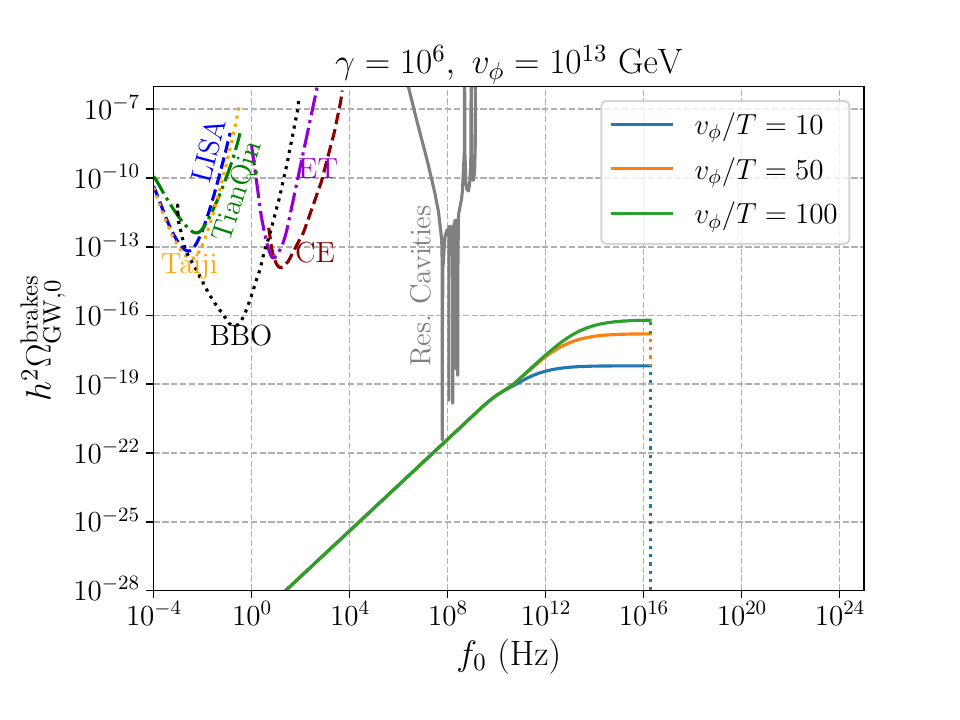}
    \caption{GW spectra generated by heavy particles traversing bubble walls during a first-order phase transition. The parameters $m$, $T$, $L_w$, and $\gamma$ are treated as free inputs. We fix the vacuum expectation value to $v_{\phi} = 10^{13}~\mathrm{GeV}$ and the bubble wall Lorentz factor to $\gamma = 10^6$. Spectra are shown for three representative cases: $v_{\phi}/T = 10$, $50$, and $100$. Other parameters are estimated as $m = v_{\phi}$ and $L_w = 1/T$. The infinite series part of the function $ I_{\rm low} $ is evaluated using its first 100 terms, which are sufficient to ensure numerical convergence.}
    \label{GWNoreheating}
\end{figure}

\begin{figure}
    \centering
    \includegraphics[width=0.88\linewidth]{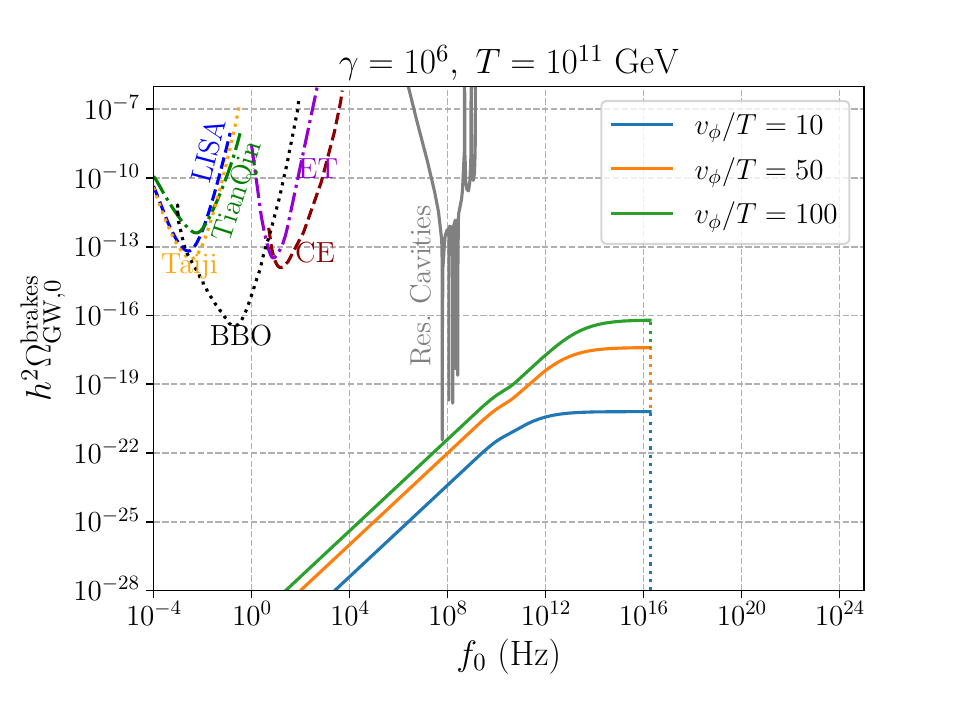}
    \caption{GW spectra generated by heavy particles interacting with bubble walls during a first-order phase transition. The parameters $m$, $v_\phi$, $L_w$, and $\gamma$ are treated as free inputs. We fix the temperature at $T = 10^{11}~\mathrm{GeV}$ and the bubble wall Lorentz factor at $\gamma = 10^6$. The figure illustrates three representative cases with $v_\phi/T = 10$, $50$, and $100$. Other parameters are estimated as $m = v_\phi$ and $L_w = 1/T$. The infinite series part of the function $ I_{\rm low} $ is evaluated using its first 100 terms, which are sufficient to ensure numerical convergence.}
    \label{GWNoreheatingga-T}
\end{figure}

\begin{figure}
    \centering
    \includegraphics[width=0.88\linewidth]{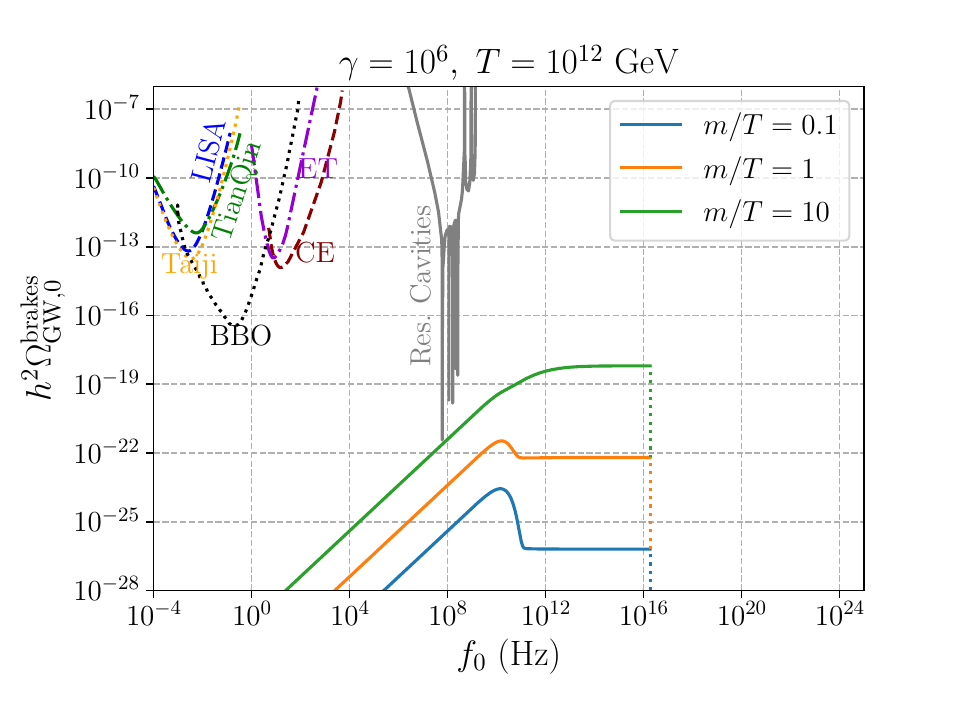}
    \caption{GW spectra generated by heavy particles interacting with bubble walls during a first-order phase transition. The parameters $m$, $v_\phi$, $L_w$, and $\gamma$ are treated as free inputs. We fix the temperature at $T = 10^{12}~\mathrm{GeV}$ and the bubble wall Lorentz factor at $\gamma = 10^6$. The figure illustrates three representative cases with $m/T = 0.1$, $1$, and $10$. The bubble wall thickness is estimated as $L_w = 1/T$. The infinite series part of the function $I_{\rm low}$ is evaluated using its first 100 terms, which are sufficient to ensure numerical convergence.}
    \label{gwmT}
\end{figure}

Figures.~\ref{GWNoreheating},~\ref{GWNoreheatingga-T} and \ref{gwmT} show the GW spectra generated by heavy particles traversing bubble walls during a first-order phase transition. The relevant parameters are treated as free inputs. We fix the vacuum expectation value at $v_{\phi} = 10^{13}~\mathrm{GeV}$ in Fig.~\ref{GWNoreheating} and the plasma temperature at $T = 10^{11}~\mathrm{GeV}$ in Fig.~\ref{GWNoreheatingga-T}, respectively, while estimating the heavy particle mass as $m = v_{\phi}$ and the bubble wall thickness as $L_w = 1/T$. GW spectra corresponding to $v_{\phi}/T = 10$, $50$, and $100$ are plotted in Figs.~\ref{GWNoreheating} and~\ref{GWNoreheatingga-T}. A strong first-order phase transition only requires \(v_\phi/T \gg 1\), so the mass can take arbitrary value in the range \(m \lesssim v_\phi\). Fig.~\ref{gwmT} shows the influence of different values of \(m/T\) on the GW spectrum for a fixed \(T=10^{12}~\mathrm{GeV}\). To obtain a numerical estimate, we evaluate the infinite series part of the function $I_{\rm low}$ using its first 100 terms, which already provide excellent convergence. This choice is retained in subsequent calculations that involve this part.
Since the Lorentz factor $\gamma$ only affects the cutoff frequency, we retain a fixed value and do not vary $\gamma$ across the plots. Sensitivity curves for proposed detectors, including LISA~\cite{LISA:2017pwj}, TianQin~\cite{TianQin:2015yph}, Taiji~\cite{Hu:2017mde},  BBO~\cite{Crowder:2005nr,Corbin:2005ny}, CE~\cite{Reitze:2019iox}, ET~\cite{Sathyaprakash:2012jk,ET:2019dnz}, and Resonant Cavities~\cite{Herman:2022fau} are also shown in Figs.~\ref{GWNoreheating} and~\ref{GWNoreheatingga-T}. 

For $\tilde{E}_k \ll T$, the function $I_{\rm low}(\tilde{E}_k)$ is predominantly governed by the second term in Eq.~\eqref{lowf}, and approaches the asymptotic form $I_{\rm low} \simeq 2\zeta_3 T^2/m^2$, where $\zeta_3 \approx 1.2$ denotes the Riemann zeta function evaluated at 3. This implies that for frequencies $f_0 \ll Ta_*/(2\pi)$, the GW amplitude scales as $h^2\Omega_{\rm GW,0}^{\rm brakes}(f_0) \propto m^2 f_0$, resulting in near-overlap of the GW amplitudes for different $v_{\phi}/T$ values in that regime when the mass is fixed, as shown in Fig.~\ref{GWNoreheating}. The critical frequency that demarcates this frequency regime is given by $f_{\rm c} = Ta_*/(4\pi)\sim 10^{10}~\mathrm{Hz}$~\footnote{For a supercooling phase transition, the calculation of redshift may need to be multiplied by a factor of \(T/T_R\), where \(T_R\) is the reheating temperature. The critical frequency becomes \(f_c \sim T/T_R \times 10^{10}~\mathrm{Hz}\).}. In the opposite limit $\tilde{E}_k \gg m$, we have $I_{\rm low} \simeq \sum_{n=1}^\infty \frac{1}{4n^2\eta} \simeq \pi^2 T/(48\tilde{E}_k)$, which is dominated by the first term in Eq.~\eqref{lowf}. Consequently, there is a nearly flat power spectrum for $f_0 \gg ma_*/(2\pi )\sim (m/T)f_c$, as shown in Figs.~\ref{GWNoreheating} and~\ref{GWNoreheatingga-T}. The GW peak frequency reaches the cutoff point $f_{\rm peak} = \gamma L_w^{-1}a_*/(2\pi)$, and the corresponding peak amplitude scales as $h^2\Omega_{\rm GW,0}^{\rm brakes}(f_{\rm peak}) \propto m^4/T^2$. Physically, the bremsstrahlung process of heavy particles predominantly radiates collinear gravitons, which therefore dominate the GW power spectrum. We consider the extreme collinear limit $k^z \gg \gamma k_\bot$, under which $\tilde{E}_k \simeq E_k/(2\gamma)$. Since the average energy of plasma particles in the bubble wall frame is $ \gamma T$, energy conservation implies $E_k < \gamma T$. Therefore, the frequency of these extremely collinear gravitons can only reach up to $f_c$; beyond this frequency, their abundance is exponentially suppressed. This behavior corresponds to the second term in Eq.~\eqref{lowf}. In contrast, the opposite extreme case $k_\bot \simeq E_k$ yields $\tilde{E}_k \simeq \gamma E_k$, and according to Eq.~\eqref{phasespace}, the frequency of non-collinear gravitons can reach $f_{\rm peak}$. This is precisely the behavior described by the first term in Eq.~\eqref{lowf}. Consequently, in the frequency range \(f_c < f_0 < f_{\rm peak}\), the power spectrum is dominated by non-collinear gravitons, whose abundance is suppressed but not exponentially. Moreover, since $\Omega_{\rm GW,0}^{\rm brakes}(f_c)/\Omega_{\rm GW,0}^{\rm brakes}(f_{\rm peak}) \simeq (48 \zeta_3/\pi^2)~T^2/m^2$, a double-peak structure emerges in the GW spectrum when $T \gtrsim m$, with an additional peak forming at $f_0 = f_c$, as shown in Fig.~\ref{gwmT}. This intermediate peak originates from contributions of light-mass particles. If the theory contains the heavy particles, i.e., $m/T \gg 1$, this peak will vanish. It is also worth emphasizing that when computing the average graviton energy in the plasma frame, one should not use the approximation $E_k \approx k^z$ (although this is valid for calculating radiation probabilities), as it would erroneously neglect contributions that are small in the wall frame but significant in the plasma frame—namely, those from non-collinear gravitons.

\begin{table}[]
    \centering
    \ccb{\begin{tabular}{cccc}
    \hline\hline
      $f_c\quad$ & $\quad\Omega_{\rm GW,0}^{\rm brakes}(f_{c})\quad$ & $f_{\rm peak}$ &  $\quad\Omega_{\rm GW,0}^{\rm brakes}(f_{\rm peak})$  \\
       \hline
         $\sim 10^{10}~\mathrm{Hz}\quad$ & $\propto m^2$ & $\quad\propto \gamma L^{-1}_w T^{-1}\quad$ & $\quad\propto m^4 T^{-2}$\\
         \hline \hline
    \end{tabular}
    \caption{Dependencies of the GW amplitude and characteristic frequencies on $\gamma$, $L_w$, $m$, and $T$. $f_c$ is the knee frequency at which the GW amplitude exhibits two distinct behavioral regimes, while $\Omega_{\rm GW,0}^{\rm brakes}(f_c)$ is the corresponding amplitude; $f_{\rm peak}$ is the ultraviolet cutoff frequency, while $\Omega_{\rm GW,0}^{\rm brakes}(f_{\rm peak})$ denotes the amplitude at this frequency.}
    \label{tab:para}}
\end{table}

In this work, we discuss a model-independent and generic mechanism for GW production via particle braking. \ccb{Table~\ref{tab:para} summarizes the dependencies of the amplitude $\Omega_{\rm GW,0}^{\rm brakes}$, the knee frequency $f_c$, and the ultraviolet cutoff frequency $f_{\rm peak}$ of such GWs on the parameters $\gamma$, $L_w$, $m$, and $T$.}
In principle, any massive Standard Model particle with arbitrary spin can radiate such GWs. However, the signal strength is tightly correlated with the particle mass. In the low-frequency regime (\(f_0 < f_c\)), \(\Omega_{\rm GW,0}^{\rm brakes}\) scales as \(m^2\), while in the high-frequency regime (\(f_c < f_0 < f_{\rm peak}\)), it scales as \(m^4\). \ccr{For this reason, standard model particles yield negligible contributions. In contrast, heavy states—such as dark matter candidates in filtered DM or Q-ball DM models~\cite{Baker:2019ndr,Chway:2019kft,Jiang:2023nkj,Witten:1984rs,Krylov:2013qe,Huang:2017kzu,Ponton:2019hux,Hong:2020est,Jiang:2023qbm,Jiang:2024zrb}—can produce substantially stronger GW signals that encode mass information. While we explicitly compute radiation from scalar particles, the method extends naturally to massive particles of arbitrary spin.
}

\section{Specific model}\label{SecModel}

In practice, the scalar particle mass $m$, plasma temperature $T$, bubble wall thickness $L_w$, and the Lorentz factor of the bubble wall $\gamma$ should be determined by the underlying phase transition model. In this section, we examine these parameters within the context of a specific model framework.

As an example, we can investigate the scale invariant $B-L$ model. We introduce one complex scalar $\Phi=(v_\phi+\phi+i\varphi)/\sqrt{2}$ that breaks the global $B-L$ symmetry. Here, $v_\phi$ is the vacuum expectation value of the phase transition field $\phi$. One extra scalar is introduced to trigger the strong first-order phase transition and produce the GW by radiating gravitons. We assume the potential of scalars has a conformal symmetry. Then it reads
\begin{equation}\label{VSp}
    V(S, \Phi)=\lambda_{s}|S|^4 + \lambda_{\phi}|\Phi|^4 +\lambda_{\phi s}|S|^2 |\Phi|^2\,\,,
\end{equation}
where $S = (s_1 + is_2)/\sqrt{2}$ is the complex scalar field that will produce braking GW. Both $s_1$ and $s_2$ could produce the GW signals discussed in Sec.~\ref{GWspectrum}. The right-handed neutrinos can couple to the field $\Phi$ through $\mathcal{L} \supset \sum_{i,j}y_{R}^{ij}\overline{\left(N_{R}^{i}\right)^c} \Phi N_R^j + \mathrm{h.c.}$.
After the $B-L$ symmetry breaking, the right-handed neutrinos acquire Majorana masses, $ M_R^{ij} = \frac{y_R^{ij}}{\sqrt{2}}v_\phi$. The neutrinos then get the small masses through the type-I seesaw mechanism
$
 m_\nu \simeq m_D^{} M_R^{-1} m_D\simeq \frac{y_D^2 v_{EW}^2}{2M_R}\,\,,
$
where $m_D = y_D v_{EW}$ is the Dirac masses after the electroweak phase transition and $v_{EW}$ is the vacuum expectation value of the Standard Model Higgs.

\subsection{The effective potential}

We get the zero-temperature potential by using the formula of Ref.~\cite{Khoze:2014xha}.
In our model, the one-loop correction from $N_R^i$ and $S$ provides the Coleman-Weinberg potential~\cite{Coleman:1973jx}
\begin{equation}\label{CW}
V_{\mathrm{CW}}^1(\phi)=\frac{1}{64 \pi^2} \sum_{k} g_k m_k^4(\phi)\left(\ln \frac{m^2_k(\phi)}{\Lambda^2}-C_k\right) \text {, }
\end{equation}
where $C_k=3 / 2$. $m_k(\phi)$ is the field-dependent mass of the particle $k$ and $g_k$ is the corresponding degree of freedom. We assume the mass matrix of right-handed neutrinos is diagonal. $\Lambda$ is a renormalization group scale. 
Substituting the field-dependent mass $m_k(\phi) \rightarrow \frac{M_k}{v_{\phi}} \phi$ into Eq.~\eqref{CW}—where $M_k$ is the measured mass of particle $k$, with
$
 M_i=y_{R, i} \frac{v_\phi}{\sqrt{2}}$ and $M_S=\sqrt{\lambda_{\phi s}}  \frac{v_\phi}{\sqrt{2}} 
$—yields the Gildener-Weinberg formula~\cite{Gildener:1976ih}. The zero-temperature potential including the tree-level potential of $\phi$ then reads
\begin{equation}\label{V0}
    V_0(\phi) = \left(A+\frac{\lambda_\phi}{4} \right)\phi^4+B \phi^4 \ln \frac{\phi^2}{\Lambda^2}\,\,,
\end{equation}
where
\begin{equation}
  \begin{aligned}
& A=\frac{1}{64 \pi^2 v_\phi^4} \sum_{k} g_k M_k^4\left(\ln \frac{M_k^2}{v_\phi^2}-C_k\right),\quad B=\frac{1}{64 \pi^2 v_\phi^4} \sum_{k} g_k M_k^4\,\,.
\end{aligned}  
\end{equation}
The minimum of the potential Eq.~\eqref{V0} provides the true vacuum expectation value, 
   $\langle\phi\rangle=v_\phi=\Lambda e^{-\left(\frac{4A+\lambda_\phi}{8B}+\frac{1}{4}\right)}$,
from which we obtain the value of the renormalization scale, $\Lambda = v_\phi e^{\left(\frac{4A+\lambda_\phi}{8B}+\frac{1}{4}\right)}$.
By substituting $\Lambda$ into Eq.~\eqref{V0} we get
\begin{equation}
V_0(\phi)=B_1 \phi^4\left(\ln \frac{\phi}{v_\phi}-\frac{1}{4}\right)\,\,, 
\end{equation}
where
\begin{equation}
   B_1=\frac{3}{2\pi^2}\left(\frac{\lambda_{\phi s}^2}{96}-\sum_i \frac{y_{R, i}^4}{96}\right)\,\,.
\end{equation}
In this work, we focus on the scenario in which scalar particles radiate gravitons. To simplify the analysis and highlight the dominant interaction, we assume $y_R^i \ll \lambda_{\phi s}$, effectively suppressing the right-handed Yukawa coupling relative to the scalar-singlet coupling. The thermal potential is given by 
\begin{equation}\label{VT}
 V_{T}(\phi,T)=\sum_{i=\mathrm{bosons}} \frac{g_i T^4}{2\pi^2}J_B\left(\frac{m_i^2(\phi)}{T^2} \right)-\sum_{i=\mathrm{fermions}} \frac{g_i T^4}{2\pi^2}J_F\left(\frac{m_i^2(\phi)}{T^2} \right)\,\,,
\end{equation}
where the thermal functions are
\begin{equation}
    J_{B,F}(x^2) = \int_{0}^{\infty} \mathrm{d}y ~ y^2 \mathrm{ln}(1\mp e^{-\sqrt{x^2+y^2}})\,\,.
\end{equation}
The Matsubara zero mode leads to infrared (IR) divergence problem. In this work, we use the Arnold-Espinosa resummation scheme~\cite{Arnold:1992rz} to alleviate the IR divergence problem.
Namely, one adds the following daisy resummation terms to the finite-temperature effective potential:
\begin{eqnarray}
    V_{\rm daisy}(\phi,T) = -\frac{T}{12\pi}\sum_{i=\mathrm{bosons}}g_i\left[\left(m_i^2(\phi)+\Pi_i(T)  \right)^{\frac{3}{2}} - m_i^3(\phi) \right],
\end{eqnarray}
where $\Pi_i(T)$ is the thermal correction for particle $i$.
Finally, the total finite-temperature effective potential becomes
\begin{eqnarray}
    V_{\rm eff}(\phi, T) = V_0(\phi) + V_T(\phi,T) + V_{\rm daisy}(\phi,T)\,\,.
\end{eqnarray}

\subsection{The phase transition dynamics}

The phase transition in the early Universe corresponds to a process of symmetry breaking, where the Universe shifts from a metastable state to a stable one. The critical temperature $T_c$ is defined as the point at which the two minima of the effective potential become degenerate, i.e., when $V_{\mathrm{eff}}(v_\phi(T_c), T_c) = V_{\mathrm{eff}}(0, T_c)$, with $v_\phi(T_c)$ being the vacuum expectation value in the true vacuum at $T = T_c$. Below $T_c$ the vacuum $\langle \phi \rangle = v_\phi(T)$ will be the global minimum. Then the transition occurs through bubble nucleation, followed by bubble growth and eventual merger. The rate at which bubbles nucleate is given by the expression
\begin{eqnarray}
\Gamma(T) \approx T^4 \left( \frac{S_3(T)}{2 \pi T} \right)^{3/2} e^{-S_3(T)/T},
\end{eqnarray}
where 
\begin{eqnarray}
S_3(T) = 4\pi \int_0^{\infty} \mathrm{d}r ~ r^2 \left[ \frac{1}{2} \left( \frac{\mathrm{d}\phi}{\mathrm{d}r} \right)^2 + V_{\mathrm{eff}}(\phi, T) \right]\label{bounce}
\end{eqnarray}
represents the action for the $\mathcal{O}(3)$-symmetric bounce solution~\cite{Salvio:2016mvj}.
The physical profile of the field $\phi$ can be determined by solving the equation of motion with the appropriate boundary conditions:
\begin{eqnarray}
\frac{\mathrm{d}^2 \phi}{\mathrm{d}r^2} + \frac{2}{r} \frac{\mathrm{d}\phi}{\mathrm{d}r} = \frac{\partial V_{\mathrm{eff}}}{\partial \phi}, \quad \phi'(0) = 0, \quad \phi(\infty) = 0.
\end{eqnarray}

The bubble nucleation begins at temperature $T_n$ which is conventionally defined by the condition $\Gamma(T_n) H^{-4}(T_n) \approx 1$, where $H(T)$ is the Hubble rate, given by
\begin{eqnarray}
H^2(T) = \frac{8 \pi}{3 M_{\mathrm{pl}}^2} \left( \frac{\pi^2}{30} g_\star T^4 + \Delta V_{\mathrm{eff}}(T) \right).
\end{eqnarray}
Here, $\Delta V_{\mathrm{eff}}(T)$ is the potential energy difference between the false and true vacua, $\Delta V_{\mathrm{eff}}(T) = V_{\mathrm{eff}}(0, T) - V_{\mathrm{eff}}(v_\phi(T), T)$. This potential difference causes bubbles to expand within the Universe, leading to a gradual reduction in the volume occupied by the false vacuum. The probability of encountering a false vacuum is given by $e^{-I(T)}$, where $I(T)$ denotes the fraction of the vacuum that has transitioned to the true vacuum, given by~\cite{Turner:1992tz, Ellis:2018mja, Megevand:2016lpr, Kobakhidze:2017mru, Ellis:2020awk, Wang:2020jrd}
\begin{eqnarray}
I(T) = \frac{4 \pi}{3} \int_T^{T_c} \mathrm{d} T^{\prime}~ \frac{\Gamma(T^{\prime})}{T^{\prime 4} H(T^{\prime})} \left[ \int_T^{T^{\prime}} \mathrm{d} \tilde{T} ~\frac{v_w}{H(\tilde{T})} \right]^3.
\end{eqnarray}
The percolation temperature $T_p$ is defined by the condition $I(T_p) = 0.34$~\cite{Turner:1992tz}, meaning that by this temperature, 34\% of the false vacuum has been converted to the true vacuum.

The strength of the phase transition at the percolation temperature is characterized by the parameter $\alpha_p$, while the inverse duration of the phase transition is given by $\beta$. They are defined as 
\begin{eqnarray}
\alpha_p = \left. \frac{\left(1 - \frac{T}{4} \frac{\partial}{\partial T}\right) \Delta V_{\mathrm{eff}}}{\pi^2 g_{*} T^4/30} \right|_{T=T_p}, \quad \beta = \left. H T \frac{\mathrm{d}}{\mathrm{d}T} \left( \frac{S_3}{T} \right) \right|_{T=T_p}.
\end{eqnarray}
The dynamics of the phase transition can be computed using CosmoTransitions~\cite{Wainwright:2011kj}. Another crucial parameter in our calculation is the width of the bubble wall $L_w$. To determine this, we use CosmoTransitions to compute the bubble wall profile, and then perform a numerical fit using the following expression for the field profile $
\phi(r) = \frac{v_\phi}{2} \left( 1 - \tanh\left( \frac{r - R}{L_w} \right) \right)$, where $R$ is the bubble radius. This fitting procedure yields a quantitative assessment of $L_w$.
If it is a supercooling first-order phase transition~\footnote{
In addition to the GWs introduced in this work, in our model, the bubble collision also provides significant GWs, as the phase transition exhibits strong supercooling and typically satisfies the runaway condition~\cite{Huber:2008hg,Caprini:2015zlo,Caprini:2019egz}. However, we observe that in our scenario, the GW signal from the braking mechanism is not stronger than that from the phase transition itself. To better distinguish between these two sources of GWs, larger Lorentz factors and lower phase transition temperatures (indicating stronger transitions) are required.},
we also need to consider the reheating effects.
By energy conservation, the reheating temperature is given by~\cite{Ellis:2019oqb}
\begin{eqnarray}
    T_R = \left(\frac{30\Delta V_{\rm eff}(T_p)}{\pi^2 g_*} \right)^{\frac{1}{4}}\times\text{min} \left(1,\frac{\Gamma_\phi}{H(T_p)}\right)^{1/2},
\end{eqnarray}
where $\Gamma_\phi$ is the decay rate of the scalar field $\phi$. Then, the GW spectrum and frequency at present become
\begin{eqnarray}
    h^2\overline{\Omega}_{\rm GW,0}^{\rm brakes}(\bar{f}_0) &=& \frac{T^4_p}{T_R^4}\left(\frac{\pi^2 g_* T^4_R/30}{\Delta V_{\rm eff}(T_p)}\right)^{\frac{4}{3}}h^2\Omega_{\rm GW,0}^{\rm brakes}(f_0)\,\, ,\\
    \bar{f}_0 &=& \frac{T_p}{T_R}\left(\frac{\pi^2 g_* T^4_R/30}{\Delta V_{\rm eff}(T_p)} \right)^{\frac{1}{3}}f_0\,\,.
\end{eqnarray}
In this work, we assume $\Gamma_\phi \gg H(T_p)$, and the relativistic degree of freedom keeps the same value $g_*$ from the
percolation temperature to the  reheating temperature.

\subsection{Prediction of bubble wall velocity}

The heavy particle braking mechanism for GW production discussed earlier relies on the bubble walls achieving highly relativistic velocities. Therefore, it is essential to ensure that the phase transition model permits such rapid bubble wall expansion. Ref.~\cite{Bodeker:2009qy} provides a criterion for determining whether bubble walls undergo runaway behavior, based on the comparison between driving and frictional forces,
\begin{eqnarray}\label{BMpanju}
    \Delta V_0 > \mathcal{F}_{\rm LO},
\end{eqnarray}
where $\Delta V_0 = \frac{B_1}{4}v_\phi^4$ denotes the vacuum energy difference between the true and false vacua at zero temperature, and $\mathcal{F}_{\rm LO}$ represents the leading-order friction arising from particle masses changing across the bubble wall. The leading-order friction is expressed as~\cite{Bodeker:2009qy}
\begin{eqnarray}\label{leadingorderF}
    \mathcal{F}_{\rm LO} = \sum_a g_a c_a \frac{\Delta m^2 T^2}{24},
\end{eqnarray}
where $g_a$ is the number of degrees of freedom of particle species $a$, $c_a = 1$ ($1/2$) for bosons (fermions), and $\Delta m$ is the corresponding mass shift. In our specific model, the dominant contribution originates from the scalar particle $S$. According to the Bodeker–Moore criterion in Eq.~\eqref{BMpanju}, the condition for runaway behavior requires $v_\phi/T > \sqrt{\lambda_{\phi s}/(6B_1)} \simeq \sqrt{32\pi^2/(3\lambda_{\phi s})}$. Some references also consider NLO friction, arising from the radiation of massive vector bosons during wall crossing. Refs.~\cite{Bodeker:2017cim,Gouttenoire:2021kjv,Azatov:2023xem} suggest $\mathcal{F}_{\rm NLO} \propto \gamma$, while Ref.~\cite{Hoche:2020ysm} proposes $\mathcal{F}_{\rm NLO} \propto \gamma^2$. In this work, since the $U(1)$ symmetry is global, we neglect NLO friction contributions.

If the theory does not include the gauge bosons or if the bubbles collide with each other before their Lorentz factor reaches the equilibrium value, the bubbles remain accelerated. By using the thin wall approximation, it is easy to get the equation of motion for the $\gamma$ factor~\cite{Ellis:2019oqb}, which is given by
\begin{equation}\label{eom}
\frac{\mathrm{d} \gamma}{\mathrm{d} R}+\frac{2 \gamma}{R}=\frac{\mathcal{F}}{\sigma},
\end{equation}
where $R$ is the bubble size, $\mathcal{F}=\Delta V_0 - \mathcal{F}_{\mathrm{LO}}$ is the pressure acting on the bubble wall, and $\sigma $ is the bubble wall tension. The critical size of the bubble $R_c$ is derived by minimizing the total energy of the bubble, $\mathcal{E}=4 \pi \gamma \sigma R^2-\frac{4 \pi}{3} R^3 \Delta V_{\rm eff}$ with $\gamma=1$, and then we have $\sigma = \Delta V_{\mathrm{eff}} R_c/2$.

After solving the equation of motion Eq.~\eqref{eom} by using the initial condition $\gamma(R_0)=1$ where $R_0$ is the initial bubble size after bubble nucleation, we get
\begin{equation}
    \gamma (R) = \frac{\mathcal{F} R}{3 \sigma}+\frac{R_0^2}{R^2}-\frac{\mathcal{P} R_0^3}{3 \sigma R^2} \approx \frac{2 R}{3 R_0}\left(\frac{\Delta V_0-\mathcal{F}_{\rm LO}}{\Delta V_{\mathrm{eff}}} \right),
\end{equation}
where we have assumed $R_c \approx R_0$ and $R \gg R_0$. The initial size $R_0$ can be expressed by the bounce action in Eq.~\eqref{bounce}, $R_0 \simeq \left(\frac{3}{2\pi}\frac{S_3}{\Delta V_{\rm eff}}\right)^{1/3}$~\cite{Ellis:2019oqb,Azatov:2019png,Levi:2022bzt}. For supercooling phase transitions, we have approximately, $\Delta V_{\mathrm{eff}} \approx \Delta V_0 \gg \mathcal{F}_{\rm LO}$.
Then the Lorentz factor at bubble collision reads $\gamma_* \simeq \frac{2}{3} \frac{R_*}{R_0}$~\cite{Azatov:2019png,Levi:2022bzt},
where $R_*$ is the bubble size at bubble collision. In Ref.~\cite{Levi:2022bzt}, it has been shown that when $\beta/H$ is large enough, the bubble size at collision can be well approximated as $R_* \approx \frac{(8\pi)^{1/3}v_w}{\beta}$. We can estimate the averaged Lorentz factor as
\begin{equation}\label{nchip}
    \bar{\gamma}= \frac{\int_{R_{0}}^{R_{*}} \mathrm{d}R ~\gamma(R)}{R_{*}-R_{0}}\approx \frac{1}{2}\gamma_*\,\, ,
\end{equation}
where we assumed $R_{*}/R_{0}= \gamma_*/\gamma_{\mathrm{min}} \gg 1$. 

\begin{figure}[!ht]
    \centering
    \includegraphics[width=0.8\linewidth]{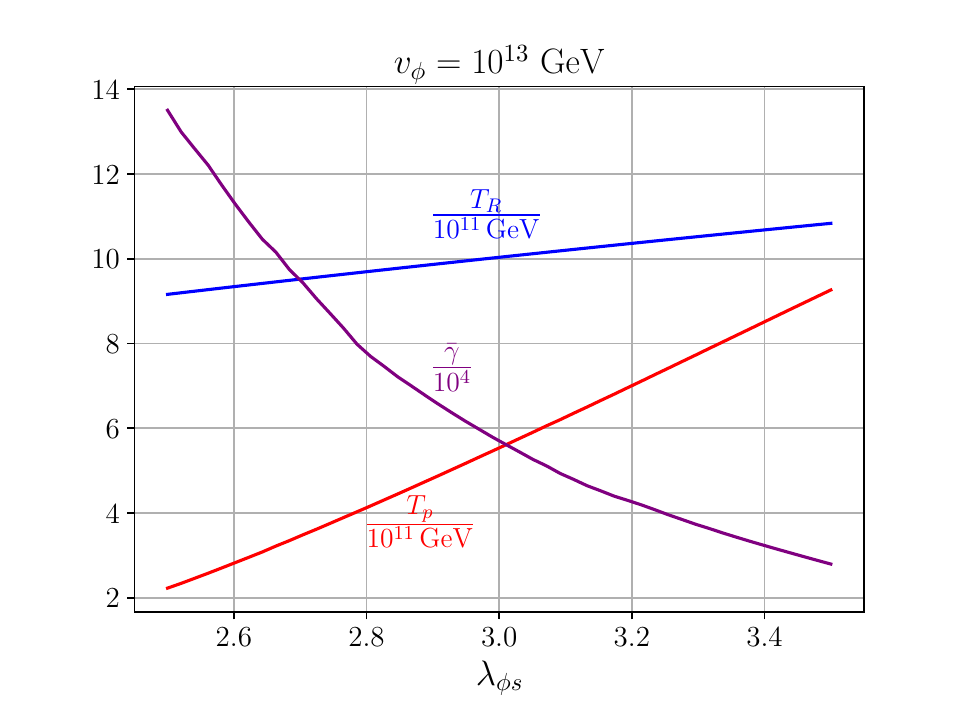}
    \caption{Dependence of phase transition dynamics parameters—the reheating temperature $T_R$, the percolation temperature $T_p$, and the average Lorentz factor of bubble walls $\bar{\gamma}$—on the portal coupling constant $\lambda_{\phi s}$. All quantities have been rescaled to dimensionless values in the range $\mathcal{O}(1)$–$\mathcal{O}(10)$ for illustrative comparison.}
    \label{parameters}
\end{figure}

In Fig.~\ref{parameters}, we present the dependence of the phase transition dynamics parameters—the percolation temperature $T_p$, the reheating temperature $T_R$, and the average Lorentz factor of the bubble walls $\bar{\gamma}$—on the portal coupling constant $\lambda_{\phi s}$. Within the examined range, the variation of $T_p$ is more pronounced than that of $T_R$, indicating that stronger phase transitions (smaller $T_p$) lead to more significant dilution effects on the GW signal due to reheating. Additionally, stronger phase transitions tend to yield larger Lorentz factors, which in turn result in higher GW peak frequencies. We observe that for $\lambda_{\phi s} > 3.5$, the bubble walls no longer satisfy the runaway condition $\Delta V_0 \gg \mathcal{F}_{\rm LO}$; therefore, we restrict our analysis to $\lambda_{\phi s} \in [2.5, 3.5]$. Moreover, we find that the bubble wall thickness approximately follows $L_w \simeq 1/T_p$, and for this reason, the variation of $L_w$ is not explicitly shown in Fig.~\ref{parameters}.

\begin{figure}
    \centering
    \includegraphics[width=0.88\linewidth]{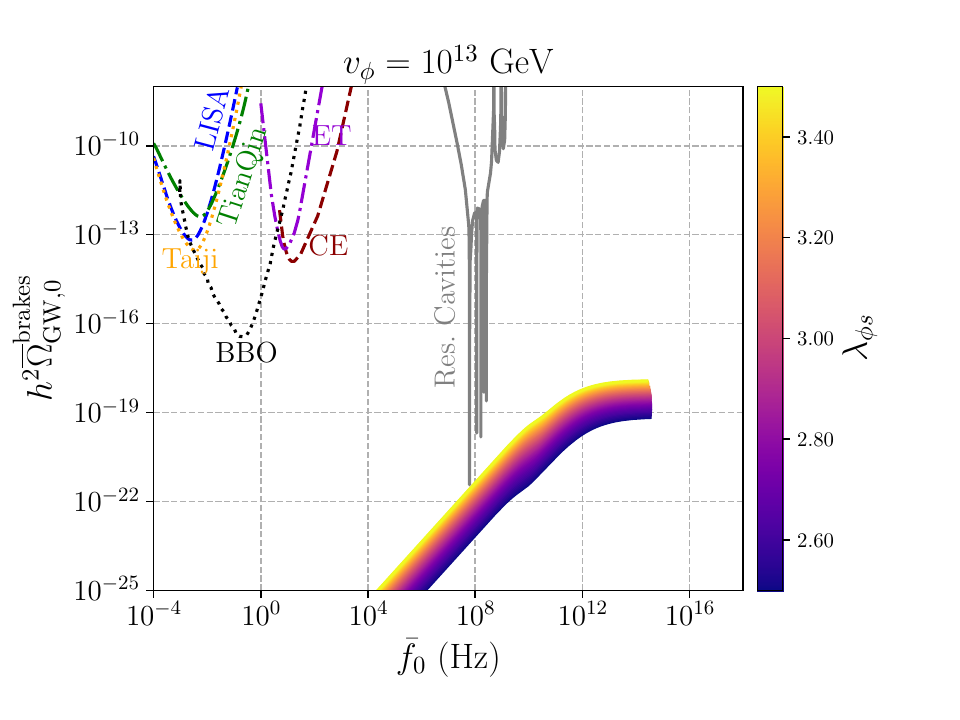}
    \caption{
GW spectra generated by heavy particles traversing bubble walls during a first-order phase transition. 
        The parameters $m$, $T$, $L_w$, and $\gamma$ are determined within the context of a $B\!-\!L$ model. 
        The vacuum expectation value is fixed at $v_\phi = 10^{13}~\mathrm{GeV}$, and the GW spectra are shown for various values of the coupling constant $\lambda_{\phi s}$, illustrating its impact on the GW spectra. The GW amplitude increases with the increasing of the portal coupling $\lambda_{\phi s}$. The infinite series part of the function $ I_{\rm low} $ is evaluated using its first 100 terms, which are sufficient to ensure numerical convergence.}
    \label{GWimshow}
\end{figure}

\subsection{Gravitational wave signals}

The resulting GW signals are presented in Fig.~\ref{GWimshow}, where the color bar illustrates the evolution of the coupling parameter $\lambda_{\phi s}$. As $\lambda_{\phi s}$ increases, the GW amplitude also becomes stronger. We analyze the behavior separately in the low-frequency regime ($\bar{f}_0<\bar{f}_c = T_p a_*/(4\pi)\propto T_p/T_R$) and high-frequency regime ($\bar{f}_c<\bar{f}_0<\bar{f}_{\rm peak}= \gamma L_w^{-1} a_*/(2\pi) \propto \gamma \left(T_p / T_R\right)$). 

In the low-frequency regime, as discussed in Sec.~\ref{GWspectrum}, the GW amplitude scales with $m_{s_1, s_2}^2 = \lambda_{\phi s} v_\phi^2 / 2$. An increase in $\lambda_{\phi s}$ leads to a reduction in the strength of the phase transition, thereby mitigating the dilution effect from reheating and enhancing the GW amplitude. In the high-frequency regime, also discussed in Sec.~\ref{GWspectrum}, the GW energy density scales as $\overline{\Omega}^{\rm brakes}_{\rm GW,0} \propto  (T_p/T_R)^4~m_{s_1, s_2}^4 / T_p^2$. For supercooled phase transitions, the reheating temperature satisfies $T_R \sim \mathcal{O}(10^{-1}) \times v_\phi$, which yields $\overline{\Omega}^{\rm brakes}_{\rm GW,0} \propto \lambda_{\phi s}^2 T_p^2$. Hence, as $\lambda_{\phi s}$ increases, the transition becomes weaker and $T_p$ increases, again leading to an enhancement in GW amplitude.

The peak frequency of the GWs is given by $\bar{f}_{\rm peak} $. Since the Lorentz factor $\gamma$ evolves faster than both $T_R$ and $T_p$, the peak frequency is slightly reduced. Although this high-frequency GW signal cannot be detected by space-based detectors such as LISA and TianQin, it may be accessible in future experiments. In particular, for sufficiently high symmetry-breaking scales, GW signals could potentially be observed via Resonant Cavities.

\section{Conclusions and discussions}\label{Summary}

In this work, we have proposed a novel mechanism for GW production during cosmological phase transitions and explicitly calculated the corresponding GW spectra through quantum field theory: the radiation of gravitons by heavy particles as they traverse bubble walls. 
Unlike conventional sources of phase transition GWs—such as bubble collisions, turbulence, or sound waves—this mechanism arises from the deceleration of massive particles interacting with the bubble wall, offering a unique probe into the microscopic properties of the early universe plasma and the dynamics of the bubble wall itself.

We have estimated the energy density of the resulting GWs and computed their power spectrum. In the ultra-relativistic limit, we derived an analytic approximation for the spectral shape, providing insight into the parametric dependence of the signal. We find that the GW power spectrum exhibits two distinct behaviors across different frequency regimes. In the low-frequency regime, the spectrum scales linearly with frequency and is proportional to the square of the mass, primarily sourced from ultra-collinear radiation emitted as particles traverse the bubble wall. In contrast, the high-frequency regime displays an approximately flat spectrum up to a cutoff frequency and the amplitude scales with the fourth power of the mass, dominated by non-collinear gravitons. For heavy particles with $ m \gg T $, the high-frequency component consistently dominates in amplitude. In this case, the GW spectrum features a single peak at the cutoff frequency $ f_{\rm peak} $, which scales with the Lorentz factor of the bubble wall $\gamma$. Notably, if all particles in the theory satisfy $ m \lesssim T $, the linearly rising low-frequency component can surpass the flat high-frequency part in amplitude. This leads to a double-peaked structure in the GW spectrum, with an additional peak emerging at the critical frequency $ f_c\sim 10^{10}~\mathrm{Hz}$ ($T_p/T_R\times10^{10}~\mathrm{Hz} $), alongside the original cutoff peak.
To illustrate the feasibility of this mechanism, we performed a concrete calculation within a specific particle physics model, demonstrating the resulting GW spectra and the dependence on model parameters.
\ccr{For these unique properties of the new GW sourcer, the contribution from Standard Model particles is negligible.
In contrast, heavy particles, such as dark matter candidates in filtered DM or Q-ball DM models~\cite{Baker:2019ndr,Chway:2019kft,Jiang:2023nkj,Witten:1984rs,Krylov:2013qe,Huang:2017kzu,Ponton:2019hux,Hong:2020est,Jiang:2023qbm,Jiang:2024zrb}, may generate significantly stronger GW signals with their mass information imprinted. In other words, future detection of such signals could offer a direct probe of DM properties, such as its mass. Although we have explicitly computed the GW radiation from scalar particles, our method can be straightforwardly generalized to massive particles of arbitrary spin. The new GW source, originating from the interplay between particle physics and cosmology, can also enrich the motivation for future high-frequency GW experiments.}

This new class of GW signals opens a complementary window into the physics of first-order phase transitions, particularly in scenarios involving heavy particle interactions. It provides a potential avenue for probing otherwise inaccessible features of the early universe, such as particle masses, interaction strengths, and bubble wall velocity.
Several important aspects remain to be explored in our future work. These include the development of more realistic particle physics models that incorporate the proposed mechanism, a systematic resummation of multi-graviton radiation contributions, and a detailed analysis of the prospects for experimental detection. Furthermore, the potential for reconstructing underlying model parameters from GW observations presents a compelling direction for future investigation. Addressing these challenges will be crucial for fully assessing the phenomenological impact and observational viability of this new GW source.

\begin{acknowledgments}

We thank Peilin Chen and Jing Yang for the helpful discussions.
This work is supported by the National Natural Science Foundation of China (NNSFC) under Grant No.12475111. and No.12205387.
    
\end{acknowledgments}

\textbf{Note added.} 
We notice that Ref.~\cite{Ai:2025fqw}, which appeared on \texttt{arXiv} one day prior to ours, investigates a closely related idea.
We find that our results are almost consistent with the renewed version of Ref.~\cite{Ai:2025fqw} in the low-frequency regime of the GW spectrum, but different in the high-frequency regime where we think the contribution from the transverse component $k_\bot$ of the gravitons could not be neglected.

\appendix

\section{Quantization of Scalar Fields in the Presence of Bubble Walls}\label{quantization}

In this appendix, we follow Ref.~\cite{Azatov:2023xem} to quantize scalar fields in the presence of bubble wall backgrounds. When bubble walls are present, the equation of motion for the scalar field $\phi$ takes the form
\begin{eqnarray}\label{kgeq}
    (\partial^2 + m^2_0 + \Delta m^2(z))\phi = 0,
\end{eqnarray}
where $m_0$ denotes the bare mass of $\phi$ in the symmetric phase, and $\Delta m^2(z)$ represents the position-dependent mass-squared correction as the field traverses the bubble wall. Assuming the mode function decomposition $\phi = e^{-i(p^0 t - p^1 x - p^2 y)}\chi(z)$, we substitute into the equation of motion to obtain a differential equation for $\chi(z)$,
\begin{eqnarray}
    [-(p^0)^2 + (p^1)^2 +(p^2)^2 + m^2_0 + \Delta m^2(z)]e^{-i(p^0 t - p^1x - p^2 y)}\chi - e^{-i(p^0 t - p^1x - p^2 y)}\chi^{\prime\prime} = 0.
\end{eqnarray}
Denote $p^z_s = \sqrt{(p^0)^2 - (p^1)^2 - (p^2)^2 - m^2_0}$, which is the longitudinal momentum of the particle in the symmetric phase, and then we have
\begin{eqnarray}
    \chi^{\prime\prime} + (p^z_s)^2\chi = \Delta m^2(z)\chi .\label{KGEQ}
\end{eqnarray}
To solve the equation, we consider two distinct regions. In the regime where the particle's wavelength is much smaller than the bubble wall width, i.e., $p^z_s \gg L_w^{-1}$, the WKB approximation becomes applicable. In this context, we explicitly restore $\hbar$ in the equation of motion, rewriting the Klein-Gordon equation as
\begin{equation}\label{eq:xh}
    \chi''(z) + \left(\frac{p^z(z)}{\hbar}\right)^2 \chi(z) = 0,
\end{equation}
where $p^z(z) = \sqrt{(p^z_s)^2 - \Delta m^2(z)}$ is a $z$-dependent effective momentum. In the absence of a bubble wall, the solution to this equation reduces to a simple plane wave, $\chi(z) \sim e^{ip^z_s z/\hbar}$. Motivated by this, we consider expanding the phase of the wavefunction in a power series of $\hbar$,
\begin{eqnarray}
    \chi(z) = e^{\frac{i}{\hbar}(S_0 + S_1\hbar+ S_2 \hbar^2 + \dots)}.
\end{eqnarray}
Substituting it into the equation of motion $\chi^{\prime \prime} + [p^z(z)/\hbar]^2\chi(z) = 0$, we have
\begin{eqnarray}
    \chi^\prime &=& \frac{i}{\hbar}\chi(z)\sum_{n=0}^\infty S_n^\prime\hbar^n = \chi(z)\sum_{n=0}^\infty iS^\prime_n\hbar^{n-1}, \\
    \chi^{\prime\prime} &=& \chi(z)\left[\sum_{n=0}^\infty iS^{\prime\prime}_n \hbar^{n-1}- \sum_{n=0}^\infty \sum_{m=0}^\infty S^{\prime}_nS^\prime_m \hbar^{m+n-2} \right],\\
      {[p^z(z)/\hbar]}^2 &=& \sum_{n=0}^\infty \sum_{m=0}^\infty S^{\prime}_nS^\prime_m \hbar^{m+n-2} -\sum_{n=0}^\infty iS^{\prime\prime}_n \hbar^{n-1} ,
\end{eqnarray}
Compare the powers of $\hbar$, we have 
\begin{eqnarray}
    \hbar^{-2} &:& [p^z(z)]^2 = (S^\prime_0)^2\\
    \hbar^{-1}&:& 2S_0^\prime S_1^\prime - iS^{\prime\prime}_0 = 0\\
    \dots && \nonumber
\end{eqnarray}
By successively solving the hierarchy of differential equations, all-order corrections to the wavefunction phase can, in principle, be obtained. For brevity, we present here the results for $S_0$ and $S_1$, respectively.
\begin{eqnarray}
    S_0 &=& \int_0^z p^z(z^\prime)dz^\prime,\\
    S_1 &=& i\ln \sqrt{\frac{p^z(z)}{p^z_s}}.
\end{eqnarray}
With this identification, we obtain the following expression:
\begin{eqnarray}
    \chi(z) = \sqrt{\frac{p^z_s}{p^z(z)}}\mathrm{exp}\left(i\int_0^z p^z(z^\prime)dz^\prime + \dots\right),
\end{eqnarray}
Here we adopt the natural unit system by setting $\hbar = 1$. Keeping only the leading-order term in the WKB expansion, we find that in the relativistic limit $p^z_s \gg \Delta m(z)$, the wavefunction admits the following approximation:
\begin{eqnarray}\label{wkbchi}
    \chi^{\rm WKB}(z) \approx e^{\pm i\int_0^z p^z(z^\prime)\mathrm{d}z^\prime},\quad p^z(z) = \sqrt{(p^z_s)^2 - \Delta m^2(z)}.
\end{eqnarray}
The other regime corresponds to particles with wavelengths larger than the bubble wall width, i.e., $p_s^z \ll L_w^{-1}$. In this case, a step-like bubble wall profile can be employed, allowing us to approximate the mass-squared variation as $\Delta m^2(z) \simeq (\tilde{m}^2 - m^2_0) \cdot \Theta(z)$, where $\Theta(z)$ denotes the Heaviside step function and $\tilde{m}$ denotes the mass of the field $\phi$ in the broken phase. Substituting into the equation of motion and solving it piecewise yields
\begin{eqnarray}
    \chi(z,p^z_s) = \left\{\begin{aligned}
        & C_1e^{ip^z_s z} + C_2e^{-ip^z_s z}, &z <0\\
        & C_3e^{ip_b^z z} + C_4 e^{-ip_b^z z}, &z\geq 0
    \end{aligned}\right.   \,\,\, ,
\end{eqnarray}
where $p_b^z = \sqrt{(p^z_s)^2 + m_0^2 -\tilde{m}^2 } = \sqrt{(p^0)^2 - (p^1)^2 -(p^2)^2 - \tilde{m}^2}$ is the longitudinal momentum of the particle in the broken phase.
To facilitate quantization, we adopt a basis consisting of ``right-moving waves" (propagating from $z = -\infty$ to $z = +\infty$, exhibiting transmission and reflection) and ``left-moving waves" (propagating from $z = +\infty$ to $z = -\infty$, also exhibiting transmission and reflection). Namely, from the general solution derived above, one can factorize the following modes:
\begin{eqnarray}\label{jidiRight}
    \chi_{R}(z,p^z_s) = N_R\left\{\begin{aligned} 
    &e^{ip^z_s z} + r_Re^{-ip^z_s z} ,&z<0,\\
    &t_Re^{ip_b^z z} , & z\geq 0,
    \end{aligned}\right.,
\end{eqnarray}
and 
\begin{eqnarray}\label{jidiLeft}
    \chi_{L}(z,p_s^z) = N_L\left\{\begin{aligned} 
    &t_Le^{-ip_s^z z} ,&z<0,\\
    & r_Le^{ip_b^z z} + e^{-ip_b^z z}, & z\geq 0,
    \end{aligned}\right. .
\end{eqnarray}
The transmission and reflection coefficients can be determined by imposing the continuity of the mode function and its derivative at the interface $z = 0$. 
\begin{eqnarray}
    &&\chi_R(0^-,p_s^z) = N_R(1+r_R) = N_Rt_R = \chi_R(0^+,p_s^z),\\ &&\chi_R^\prime(0^-,p_s^z) = N_R(ip_s^z - ir_Rp_s^z) = N_Rit_Rp_b^z = \chi_R^\prime(0^+,p_s^z).
\end{eqnarray}
These boundary conditions ensure that the physical solution across the potential discontinuity remains well-behaved and conserves probability flux. Thus, one can get the solution
\begin{eqnarray}
    r_R = \frac{p^z_s - p_b^z}{p_s^z + p_b^z},\quad t_R = \frac{2p_s^z}{p_s^z + p_b^z}.
\end{eqnarray}
Similarly, we can also obtain
\begin{eqnarray}
    r_L = \frac{p_b^z -p_s^z}{p_s^z +p_b^z},\quad t_L = \frac{2p_b^z}{p_s^z +p_b^z}.
\end{eqnarray}
The normalization coefficients $N_R$ and $N_L$ are determined by imposing the normalization condition on the basis states. It is important to note that, by definition, both $p_s^z$ and $p_b^z$ are constrained to be positive. This choice simplifies the analysis by ensuring that the momentum components along the $z$-axis remain well-defined and physically consistent with the chosen direction of propagation.
Then, we can have
\begin{eqnarray}
    &&\int_{-\infty}^\infty\mathrm{d}z~\chi_R(z,p^z_s)\chi_R^*(z,q^z_s) \nonumber\\
    &=& N_R^2\int_{-\infty}^0 \mathrm{d}z~ (e^{ip^z_s z} +r_{R,p}e^{-ip^z_s z})(e^{-iq^z_s z} +r_{R,q}e^{iq^z_s z}) 
    + N_R^2\int_0^\infty \mathrm{d}z~t_{R,p}t_{R,q}e^{i(p_b^z -q_b^z)z  }\nonumber\\
    &=& N_R^2 \left(\mathrm{PV}\left[\frac{1}{i(p_s^z - q_s^z)}\right] + \pi\delta(p_s^z-q_s^z) +\mathrm{PV}\left[\frac{r_{R,q} - r_{R,p}}{i(p_s^z +q_s^z)}\right]+\mathrm{PV}\left[\frac{r_{R,p}r_{R,q}}{-i(p_s^z - q_s^z )}\right]\right. \nonumber\\
    &&\left.+  r_{R,p}r_{R,q}\pi\delta(q_s^z - p_s^z) - \mathrm{PV}\left[\frac{t_{R,p}t_{R,q}}{i(p_b^z - q_b^z)}\right] + t_{R,p}t_{R,q}\pi\delta(p_b^z - q_b^z)   \right)\nonumber\\
    &=&N_R^2 2\pi \delta(p_s^z-q_s^z),  \quad \Rightarrow \quad N_R =1.\label{zjgyijj=s}
\end{eqnarray}
The above integration has used the identity
\begin{eqnarray}
    \int_{-\infty}^0\mathrm{d}z~e^{i\beta z} = \mathrm{PV}\left(\frac{1}{i\beta} \right) + \pi\delta(\beta).
\end{eqnarray}
To deal with the Cauchy principal value (PV), one can use
 $(p_s^z)^2 - (q_s^z)^2 = (p_b^z)^2 - (q_b^z)^2$ and $t_R = 1 + r_R$. 

Applying the same procedure to $\chi_L$ yields the normalization condition $N_L^2 = p_s^z / p_b^z$. Utilizing these results, we further derive the following inner product relation:
\begin{eqnarray}
    \int_{-\infty}^{\infty} \mathrm{d}z~\chi_I(z,p^z_s)\chi^*_J(z,q^z_s) &=& 2\pi \delta_{IJ}\delta(p^z_s -q_s^z), \quad I,J\in \{R,L\},\label{zhenjiaogx}\\
    \int_{-\infty}^\infty \mathrm{d}z~\chi_R(z,p^z_s)\chi_R(z,q^z_s)&=&-\int_{-\infty}^\infty \mathrm{d}z~\chi_L(z,p^z_s)\chi_L(z,q^z_s) = 2\pi \frac{p_s^z - p_b^z}{p_s^z + p_b^z}\delta(p_s^z -q_s^z),\,\, \,\label{neiji2}\\
    \int_{-\infty}^\infty \mathrm{d}z~\chi_R(z,p^z_s)\chi_L(z,q^z_s)&=& 4\pi \frac{\sqrt{p_s^z p_b^z}}{p_s^z + p_b^z}\delta(p_s^z -q_s^z). \label{neiji}
\end{eqnarray}

In fact, within the regime where the WKB approximation is valid, the basis functions take a form analogous to those in Eq.~\eqref{jidiRight} and Eq.~\eqref{jidiLeft}, though the transmission and reflection coefficients may become $z$-dependent and asymptote to their step-wall values as $z \rightarrow \pm \infty$. 
As indicated by Eq.~\eqref{wkbchi}, given that $p_s^z \sim p_b^z \gg L_w^{-1} \sim \sqrt{\tilde{m}^2 - m_0^2}$, the momentum $p^z(z)$ can be expanded near $z = \pm \infty$ using a Taylor series
\begin{eqnarray}
p^z(z) = p_s^z + \mathcal{O}\left[\frac{\Delta m^2(z)}{(p_s^z)^2}\right] = p_b^z + \mathcal{O}\left[\frac{\Delta m^2(z)}{(p_b^z)^2}\right].
\end{eqnarray}
This leads to the approximate WKB solution,
\begin{eqnarray}
    \chi^{\rm WKB}(z,p^z_s) \approx \left\{\begin{aligned}
        &\xi_{<0}(z) e^{\pm i p^z_s z}, &z<0,\\
        &\xi_{>0}(z) e^{\pm i p_b^z z }, & z\geq 0,
    \end{aligned}\right.
\end{eqnarray}
The ``$+$'' and ``$-$'' signs correspond to right-moving and left-moving waves, respectively. Compared to Eq.~\eqref{jidiRight} and Eq.~\eqref{jidiLeft}, this solution lacks reflected wave contributions. In the limit $p_s^z \sim p_b^z$, we find $r_R \sim r_L \sim 0$ and $t_R \sim t_L \sim 1$, indicating that reflection effects are negligible. Therefore, whether in the WKB region or in the step-wall region, the complete basis given in Eq.~\eqref{jidiRight} and Eq.~\eqref{jidiLeft} remains applicable, though the coefficients must be computed differently in each case. Therefore, by incorporating the transverse plane wave components, we obtain the  ``plane wave solution'' that satisfies the Klein–Gordon equation,
\begin{eqnarray}
   && \phi_R(p) = e^{-i p_n x^n}\chi_R(z,p_s^z),\quad p_nx^n = p^0 t - \vec{p}_\bot \cdot \vec{x}_\bot ,\  p^0 > m_0,\\
   && \phi_L(p) = e^{-i p_n x^n}\chi_L(z,p_s^z),\quad p_nx^n = p^0 t - \vec{p}_\bot \cdot \vec{x}_\bot ,\  p^0 > \tilde{m} .
\end{eqnarray}

Then the scalar field $\phi(x)$ can then be expanded in terms of these basis functions, 
providing a complete representation of the field within the given framework.
\begin{eqnarray}
    \phi(x) = \sum_{I=R,L}\int\frac{\mathrm{d}^3p}{(2\pi)^3\sqrt{2p^0}}\left(a_{I,p}\phi_I(p)  + a^\dag_{I,p}\phi^*_I(p)   \right),
\end{eqnarray}
where the operators $a_{I,p}$ and $a_{I,p}^\dag$ correspond to the annihilation and creation operators, respectively. The conjugate momentum density is given by
\begin{eqnarray}
    \pi(x) \equiv \partial_t \phi(x) = \sum_{I=R,L}\int\frac{\mathrm{d}^3p}{(2\pi)^3\sqrt{2p^0}}\left(-ip^0 a_{I,p}\phi_I(p)  + ip^0a^\dag_{I,p}\phi^*_I(p)   \right).
\end{eqnarray}
It is important to note that $\mathrm{d}^3p = \mathrm{d}p^z_s\mathrm{d}^2p_\bot$ and the integration domain for $p^z_s$ is restricted to the interval $[0, \infty)$. When $\sqrt{(p^0)^2 - p_\bot^2} < \tilde{m}$, the mode function $\phi_L(p)$ vanishes identically, i.e., $\phi_L(p) \equiv 0$. By employing the orthogonality relations given in Eq.~\eqref{zhenjiaogx}, Eq.~\eqref{neiji2}, and Eq.~\eqref{neiji}, we can derive the explicit expression for the creation and annihilation operators.
\begin{eqnarray}
    \int \mathrm{d}^3x ~ \phi^*_I(p)\phi(x) &=& \sum_{J=R,L}\int \mathrm{d}z\int\frac{\mathrm{d}^3q}{(2\pi)^3\sqrt{2q^0}} 
\left(a_{J,q}\chi_J(z,q_s^z)\chi^*_I(z,p_s^z)e^{i(p^0 - q^0) t} \right.\nonumber\\
&& \left.\times(2\pi)^2\delta^{(2)}(\vec{q}_\bot - \vec{p}_\bot)+ a^\dag_{J,q}\chi^*_J(z, q^z_s)\chi^*_I(z,p^z_s)e^{i(p^0 + q^0) t} (2\pi)^2\delta^{(2)}(\vec{q}_\bot + \vec{p}_\bot)\right)\nonumber\\
    &=& \int\frac{\mathrm{d}^3q}{(2\pi)^3\sqrt{2q^0}} (2\pi)^3\delta(q_s^z - p_s^z)
   \left[a_{I,q}e^{i(p^0 - q^0) t}\delta^{(2)}(\vec{q}_\bot - \vec{p}_\bot)\right. \nonumber\\
   &&+ \left.\left(2\frac{\sqrt{q_s^z q_b^z}}{q_s^z +q_b^z} + C_{I} \frac{q_s^z - q_b^z}{q_s^z +q_b^z}\right)a^\dag_{I,q}e^{i(p^0 + q^0) t}\delta^{(2)}(\vec{q}_\bot + \vec{p}_\bot) \right]\nonumber\\
    &=& \frac{1}{\sqrt{2p^0}}\left[a_{I,p} + \left(2\frac{\sqrt{q_s^z q_b^z}}{q_s^z +q_b^z} + C_I \frac{q_s^z - q_b^z}{q_s^z +q_b^z}\right)a^\dag_{I,-p_\bot}e^{2ip^0 t} \right],\\
     \int \mathrm{d}^3x ~ \phi^*_I(p)\pi(x) &=& \frac{-ip^0}{\sqrt{2p^0}}\left[ a_{I,k} - \left(2\frac{\sqrt{q_s^z q_b^z}}{q_s^z +q_b^z} + C_I \frac{q_s^z - q_b^z}{q_s^z +q_b^z}\right)a^\dag_{I,-p_\bot}e^{2ip^0 t} \right],
\end{eqnarray}
where $C_R = 1$ and $C_L=-1$.
Thus, one can get
\begin{eqnarray}
 a_{I,p} = \int\frac{\mathrm{d}^3 x}{\sqrt{2p^0}}\phi^*_I(p)[p^0\phi(x) + i\pi(x)].
\end{eqnarray}
By applying the equal-time commutation relations between the field $\phi(x)$ and its conjugate momentum $\pi(x)$, we obtain the following result:
\begin{eqnarray}
    \left[a_{I,p},a_{J,q}^\dag  \right] &=& (2\pi)^3\delta^{(2)} (\vec{p}_\bot - \vec{q}_\bot)\delta(p^z_s - q^z_s)\delta_{IJ},\\
    \Big[a_{I,p},a_{J,q}  \Big] &=& \left[a^\dag_{I,p},a^\dag_{J,q}  \right] = 0,\quad I,J\in \{R,L\}.
\end{eqnarray}
The single particle states are defined by
\begin{eqnarray}
    \ket{p^R} &\equiv& \sqrt{2p^0} a^\dag_{R,p}\ket{0},\\
    \ket{p^L} &\equiv& \sqrt{2p^0} a^\dag_{L,p}\ket{0}.
\end{eqnarray}
Notice that this particle state corresponds to the incident state at $t\rightarrow-\infty$, while the outgoing state at $t\rightarrow + \infty$ is superposition state of transmission state and reflection state. By using the time reversal, we can get another set of orthogonal bases,
\begin{eqnarray}
    \phi^{\rm out}_{L}(p) = e^{-ip_n x^n}\zeta_L(z,p^z_s) &=& e^{-ip_n x^n}\chi^*_R(z,p^z_s) \nonumber\\
    &=& e^{-ip_n x^n}\left( r^*_{R,p}\chi_{R}(z,p^z_s) + t^*_{R,p}\sqrt{\frac{p_b^z}{p^z_s}}\chi_L(z,p_s^z) \right),\quad\\
     \phi^{\rm out}_{R}(p) = e^{-ip_n x^n}\zeta_R(z,p^z_s) &=& e^{-ip_n x^n}\chi^*_L(z,p_s^z) \nonumber\\
     &=& e^{-ip_n x^n}\left( r^*_{L,p}\chi_{L}(z,p^z_s) + t^*_{L,p}\sqrt{\frac{p^z_s}{p_b^z}}\chi_R(z,p_s^z) \right).\quad
\end{eqnarray}
These bases correspond to the outgoing particle states,
\begin{eqnarray}
    \ket{p^{L,\mathrm{out}}} &=& r^*_{R,p}\ket{p^R} + t^*_{R,p}\sqrt{\frac{p_b^z}{p_s^z}}\ket{p^L},\\
     \ket{p^{R,\mathrm{out}}} &=& t^*_{L,p}\sqrt{\frac{p_s^z}{p_b^z}}\ket{p^R} +  r^*_{L,p}\ket{p^L}\,\,.
\end{eqnarray}
And the form and the commutation relation of creation and annihilation operators are similar with that of the incident state. More details about particle states can be seen in Ref.~\cite{Azatov:2023xem}.

\section{Splitting probability}\label{probability}
For the process $a(p_a) \rightarrow b_1(p_1) b_2(p_2)\dots b_n(p_n)$, the splitting probability after integration over the final states reads
\begin{eqnarray}
     \int \mathrm{d}P_{1\rightarrow n} &\equiv& \left(\prod_{j=1}^n\tilde{V}\int \frac{\mathrm{d}^3 p_j}{(2\pi)^3}\right) \frac{|\braket{\vec{p}_1,\dots,\vec{p}_n|\mathcal{T}|\phi_a}|^2}{\braket{\phi_a|\phi_a}\prod_{j=1}^n\braket{\vec{p}_j|\vec{p}_j}},
\end{eqnarray}
 where $\tilde{V}$ denotes the volume of the spatial integration range. The normalized state $\ket{\phi_a}$ is defined as
\begin{eqnarray}\label{braphi}
    \ket{\phi_a} \equiv \int \frac{\mathrm{d}^3 p_a}{(2\pi)^3}\frac{\phi(\vec{p}_a)}{2E_a}\ket{\vec{p}_a},\quad \int \frac{\mathrm{d}^3 p_a}{(2\pi)^3}\frac{|\phi(\vec{p}_a)|^2}{2E_a}=1,\quad \ket{\vec{p}_i} = \sqrt{2E_i}a^\dag_{i}\ket{0}.
\end{eqnarray}
It has a normalized inner product
\begin{eqnarray}
    \braket{\phi_a|\phi_a} = 1,\quad \braket{\vec{p}_j|\vec{p}_j} = 2E_{\vec{p}_j}(2\pi)^3\delta^{(3)}(\vec{p}_j-\vec{p}_j) = 2E_{\vec{p}_j}\int \mathrm{d}^3x ~e^{i(\vec{p}_j-\vec{p}_j)\cdot \vec{x}} =  2E_{\vec{p}_j} \tilde{V} .
\end{eqnarray}

Using the quantization results in Appendix~\ref{quantization}, we can calculate the interaction matrix element of the splitting process. In this work, we are mainly concerned with the tree-level contribution of the process where a scalar particle acquires mass while crossing the bubble wall and radiates a graviton, $s(p_a)\rightarrow s(p_b)g(k)$. Therefore, in this appendix, we mainly present the calculation of the $S$-matrix for such a process. The interaction Hamiltonian density involved in this process can be expressed as
\begin{eqnarray}
    \mathcal{H}_{\rm int} &=& \sum_{I,J=R,L}\int\frac{\mathrm{d}^3p_a}{(2\pi)^3}\int\frac{\mathrm{d}^3p_b}{(2\pi)^3}\int\frac{\mathrm{d}^3k}{(2\pi)^3} ~ \big[V(z)a_{k}\chi(z,k^z)+ V^\dag(z)a_{k}^\dag\chi^* (z,k^z)\big] e^{-i(E_k t -\vec{k}_{\bot}\cdot \vec{r}_{\bot})} \nonumber\\
    && \times \big[a_{I,p_a}\chi_{I}(z,p_a^z)+ a_{I,p_a}^\dag\chi_I^* (z,p_a^z)\big] \big[a_{J,p_b}\zeta_{J}(z,p_b^z)+ a_{J,p_b}^\dag \zeta_J^* (z,p_b^z)\big] \nonumber\\
    &&\times e^{i (E_a t - \vec{p}_{a,\bot}\cdot \vec{r}_{\bot})} e^{-i(E_b t - \vec{p}_{b,\bot}\cdot \vec{r}_{\bot})} \quad ,
\end{eqnarray}
where the vertex function $V(z)$ may include momentum operators and coupling constants. The two outgoing final states of the scalar particle (left-moving and right-moving) correspond to different creation and annihilation operators, while the incoming initial state is right-moving. Therefore, the interaction matrix element becomes
\begin{eqnarray}
    \bra{\vec{p}_b^{I,\mathrm{out}},\vec{k}} \mathcal{T} \ket{\vec{p}_a^R} &=& \int \mathrm{d}^4 x  \bra{\vec{p}_b^{I,\mathrm{out}},\vec{k}}\mathcal{H}_{\rm int}\ket{\vec{p}_a^R} \nonumber\\
    &=& \int \mathrm{d}z\int\frac{\mathrm{d}^3p_a^\prime}{(2\pi)^3}\int\frac{\mathrm{d}^3p_b^\prime}{(2\pi)^3}\int\frac{\mathrm{d}^3k^\prime}{(2\pi)^3} ~V^\dag(z)\chi_{R}(z,p_a^{\prime z})\zeta^*_{I}(z,p_b^{ \prime  z}) \chi^*(z,k^{ \prime z}) \nonumber\\
    &&\times (2\pi)^3 \delta(E_a^\prime - E^\prime_b - E_k^\prime)\delta^{(2)}(\vec{p}_{a,\bot}^\prime -\vec{p}_{b,\bot}^\prime - \vec{k}_{\bot}^\prime)\bra{\vec{p}_b^{I,\mathrm{out}},\vec{k}}a^\dag_k a^\dag_{I,b} a_{R,a}\ket{\vec{p}_a^R}\nonumber\\
    &=& (2\pi)^3 \delta\left(\sum E\right) \delta^{(2)}\left(\sum \vec{p}_{\bot}\right)\mathcal{M}_I,
\end{eqnarray}
where the matrix element
\begin{eqnarray}\label{juzhenyuan}
    \mathcal{M}_I = \int_{-\infty}^{+\infty} \mathrm{d}z~ V^\dag(z)\chi_{R}(z,p_a^z)\zeta^*_{I}(z,p_b^z) \chi^*(z,k^z).
\end{eqnarray}
Thus, the probability of the $s\rightarrow sg$ process is written as
\begin{eqnarray}
    \int \mathrm{d}P_{s\rightarrow sg} &=& \int\frac{\mathrm{d}^3p_a^{\prime}}{(2\pi)^3 2E_a^\prime} \int\frac{\mathrm{d}^3 p_b}{(2\pi)^3 2E_b} \int\frac{\mathrm{d}^3 k}{(2\pi)^32E_k} \int\frac{\mathrm{d}^3p_a^{\prime \prime}}{(2\pi)^3 2E_a^{\prime \prime}} \phi^*(\vec{p}_a^{\prime})\phi(\vec{p}_a^{\prime\prime}) \nonumber\\
    && \times \sum_{I=R,L} \bra{\vec{p}_a^{\prime R}} \mathcal{T}\ket{\vec{p}_b^{I,\mathrm{out}},\vec{k}} \bra{\vec{p}_b^{I,\mathrm{out}},\vec{k} }\mathcal{T}\ket{\vec{p}_a^{\prime\prime R}}\nonumber\\
    &=& \int\frac{\mathrm{d}^3p_a^{\prime}}{(2\pi)^3 2E_a^\prime} \int\frac{\mathrm{d}^3 p_b}{(2\pi)^3 2E_b} \int\frac{\mathrm{d}^3 k}{(2\pi)^32E_k} \int\frac{\mathrm{d}^3p_a^{\prime \prime}}{(2\pi)^3 2E_a^{\prime \prime}} \phi^*(\vec{p}_a^{\prime})\phi(\vec{p}_a^{\prime\prime}) (2\pi)^6 \delta\left(\sum E^{\prime}\right)  \nonumber\\
    && \times \delta^{(2)}\left(\sum \vec{p}^{\prime}_{\bot}\right)  \delta\left(\sum E^{\prime\prime}\right)\delta^{(2)}\left(\sum \vec{p}^{\prime\prime}_{\bot}\right) \sum_{I=R,L}\mathcal{M}^*_I(p_a^{\prime\prime}) \mathcal{M}_I(p_a^\prime), 
\end{eqnarray}
where $(2\pi)^2\delta^{(2)}(\sum \vec{p}^{\prime\prime}_{\bot})$ can be eliminated by integrating out the transverse momentum of $\vec{p}_a^{\prime\prime}$, and $2\pi\delta(\sum E^{\prime\prime})$ can be eliminated by integrating out the remaining components of $\vec{p}^{\prime\prime}_a$, yielding an additional factor $E^{\prime\prime}_a/p^{ \prime\prime z}_a$. Combined with the other two $\delta$ functions, it can be seen that $\vec{p}^{\prime\prime}_a$ can be replaced by $\vec{p}^{\prime}_a$. Thus, the splitting probability becomes
\begin{eqnarray}
    \int \mathrm{d}P_{s\rightarrow sg} &=& \int\frac{\mathrm{d}^3 p_b}{(2\pi)^3 2E_b} \int\frac{\mathrm{d}^3 k}{(2\pi)^32E_k} \int\frac{\mathrm{d}^3 p_a^\prime}{(2\pi)^3}\frac{|\phi(\vec{p}_a^\prime)|^2}{2E^\prime_a}\frac{1}{2p_a^{\prime z}} \nonumber\\
    && \times (2\pi)^3\delta^{(2)}\left(\sum \vec{p}^\prime_{\bot} \right)  \delta\left(\sum E^\prime \right)\left(|\mathcal{M}_R|^2 + |\mathcal{M}_L|^2\right).
\end{eqnarray}
Assuming that $\phi(\vec{p})$ is highly localized around $\vec{p} = \vec{p}_a$, in conjunction with the right-hand side of Eq.~\eqref{braphi}, the function $\frac{|\phi(\vec{p})|^2}{2E}$ can be considered approximately as a $(2\pi)^3\delta^{(3)}(\vec{p} - \vec{p}_a)$ function. Therefore, we have finally
\begin{eqnarray}
    \int \mathrm{d}P_{s\rightarrow sg} &=& \int\frac{\mathrm{d}^3 p_b}{(2\pi)^3 2E_b} \int\frac{\mathrm{d}^3 k}{(2\pi)^32E_k} \int\frac{\mathrm{d}^3 p_a^\prime}{(2\pi)^3 2p^{\prime z}_a}(2\pi)^3\delta^{(3)}(\vec{p}^\prime_a - \vec{p}_a) \nonumber\\
    && \times (2\pi)^3\delta^{(2)}\left(\sum \vec{p}^\prime_{\bot} \right)  \delta\left(\sum E^\prime \right)\left(|\mathcal{M}_R|^2 + |\mathcal{M}_L|^2\right)\nonumber\\
    &=& \int\frac{\mathrm{d}^3 p_b}{(2\pi)^3 2E_b} \int\frac{\mathrm{d}^3 k}{(2\pi)^32E_k}\frac{1}{2p_{a,s}^z} (2\pi)^3 \delta^{(2)}\left(\sum \vec{p}_{\bot} \right)  \delta\left(\sum E \right)\nonumber\\
    &&\times \left(|\mathcal{M}_R|^2 + |\mathcal{M}_L|^2\right).
\end{eqnarray}
It should be emphasized again here that $\mathrm{d}^3p_b = \mathrm{d}p^z_{b,s}\mathrm{d^2}p_{b,\bot}$ with $p^z_{b,s} \in [0, \infty)$. Both $p^z_{b,s}$ and $p^z_{a,s}$ are the longitudinal momentum values in the symmetric phase.

\ccb{\section{An example of the transition amplitude suppression region $\Delta p^z_b L_w \gg 1$}\label{non-adiabaticity}
In Sec.~\ref{probabilitySec}, we discussed that when $\Delta p^z_bL_w \gg 1$ the transition amplitude is strongly suppressed. In this appendix, we calculate the transition amplitude using a specific bubble wall profile to check this condition (see also Appendix H of~\cite{Azatov:2023xem} for a similar discussion). The matrix element of transition has the following general form
\begin{equation}
    \mathcal{M}^{\rm WKB}= \int_{-\infty}^{\infty}\mathrm{d}z~V(z)e^{i\int_0^z\mathrm{d}z^\prime\Delta p^z(z^\prime)}.\label{MWKBC1}
\end{equation}
For simplicity, we can typically choose the function $\Delta p^z(z)$ as
\begin{equation}
    \Delta p^z(z) = \Delta p^z_0 + \varepsilon\tanh \frac{z}{L_w},
\end{equation}
where $\Delta p^z_0 = (\Delta p^z_b + \Delta p^z_s)/2$ and $\varepsilon = (\Delta p^z_b - \Delta p^z_s)/2$. It is noted that the vertex used in Eq.~\eqref{vertexfunc} at leading order is
\begin{equation}
    V(z) \simeq V_0 = -\frac{i\kappa k^2_\bot}{2x^2 E_a} ,
\end{equation}
which is independent of $z$. Then the matrix element becomes
\begin{equation}
    \mathcal{M}^{\rm WKB} = V_0 \int_{-\infty}^{\infty} \mathrm{d}z ~\exp\left[i\Delta p^z_0 z + i\varepsilon f(z)\right],
\end{equation}
where
\begin{equation}
    f(z) = L_w \ln \cosh \frac{z}{L_w}.
\end{equation}
Finally, we can obtain an analytical expression
\begin{align}
|\mathcal{M}^{\rm WKB}|^2 &= \frac{\pi \varepsilon L_w |V_0|^2}{(\Delta p^z_0)^2 - \varepsilon^2} \times \frac{\sinh\left( \pi \epsilon L_w  \right)}{\sinh\left( \pi (\Delta p^z_0 - \varepsilon) L_w / 2 \right) \sinh\left( \pi (\Delta p^z_0 + \varepsilon) L_w / 2 \right)} \\
&\approx \frac{\pi \varepsilon L_w |V_0|^2}{(\Delta p^z_0)^2} \times \frac{\sinh\left( \pi \varepsilon L_w  \right)}{\sinh^2\left( \pi \Delta p^z_0 L_w / 2 \right)}.
\end{align}
Under the soft-collinear and the ultra-relativistic approximations, we have
\begin{align}
    \varepsilon \simeq \frac{xm^2}{2E_a}\ll L_w^{-1}.
\end{align}
For $\Delta p_0^z L_w\gg 1$ (equivalent to $\Delta p_b^z L_w\gg 1$), we derive
\begin{equation}
    |\mathcal{M}^{\rm WKB}|^2 \approx \frac{\pi \varepsilon L_w |V_0|^2}{ (\Delta p^z_0)^2} \times \pi \varepsilon L_w e^{-\pi \Delta p^z_0 L_w} = \frac{(\pi \varepsilon L_w)^2 |V_0|^2}{ (\Delta p^z_0)^2} \times e^{-\pi \Delta p^z_0 L_w},
\end{equation}
which is exponentially suppressed by the factor $e^{-\pi \Delta p^z_0 L_w}$.
}

\bibliography{ref}

\end{document}